\documentclass[final,3p,times]{elsarticle}
\usepackage{CJK}
\usepackage{graphicx} %We can use any other package if it is necessary
\usepackage{amssymb}
\usepackage{amsmath}
\usepackage{subfigure}
\usepackage{graphicx}
\usepackage{epstopdf}
\usepackage{array}
\usepackage{bibentry}
\usepackage{mathrsfs}
\usepackage{color}
\biboptions{sort&compress}
\usepackage{breqn}
\usepackage{lineno,hyperref}
\DeclareMathOperator{\sech}{sech}
\newtheorem{thm}{Theorem}[section]

\newtheorem{prop}[thm]{Proposition}
\newtheorem{cor}[thm]{Corollary}

\allowdisplaybreaks[4]

\journal{}

\begin{document}

\begin{frontmatter}
\title{The dynamic of the positons for the reverse
	space-time nonlocal short pulse equation}

\author{Jiaqing Shan}
\author{\corref{cor1}Maohua Li}
\address{School of Mathematics and Statistics, Ningbo University,
	Ningbo, 315211, P.\ R.\ China}

\cortext[cor1]{Corresponding author: limaohua@nbu.edu.cn;limaohua@mail.ustc.edu.cn}

\begin{abstract}
In this paper, the Darboux transformation (DT) of the reverse space-time (RST) nonlocal short pulse equation is constructed by a hodograph transformation and the eigenfunctions of its Lax pair. The multi-soliton solutions of the RST nonlocal short pulse equation are produced through the DT, which can be expressed in terms of determinant representation. By taking different values of eigenvalues, bounded soliton solutions and unbounded soliton solutions can be obtained. In addition, based on the degenerate Darboux transformation, the $N$-positon solutions of the RST nonlocal short pulse equation are computed from the determinant expression of the multi-soliton solution. Furthermore, different kinds of mixed solutions are also presented, and the interaction properties between positons and solitons are investigated.
\end{abstract}
\begin{keyword}
	Reverse space-time nonlocal short pulse equation, Darboux transformation, Positon, Mixed solution,  Interaction.
\end{keyword}
\end{frontmatter}

\section{Introduction}
\numberwithin{equation}{section}
The short pulse equation (SPE) is introduced by Sch\"{a}fer and Wayneo\cite{Schafer2004,Schafer2005} in order to describe the propagation of ultra-short optical pulses in quartz fibers. Its expression is as follows 
\begin{align}\label{a1}
	q_{xt}=2q+\frac{1}{3}(q^3)_{xx}.
\end{align}

Initialiy, it appeared as an integrable differential equation to describe the pseudosphere surfaces\cite{Rabelo1989}. It has been proven that the SP equation is integrable and it possesses infinite conservation law\cite{Feng2015,Zhang2015}, the bi-Hamiltonian structure\cite{Brunelli2005,Brunelli2006} and a Lax pair of the Wadati-Konno-Ichikawa (WKI) type\cite{Sakovich2005}. The hodograph transformation between the SP equation and the sine-Gorden equation has been established in\cite{Sakovich2005, Matsuno2007}. Furtherly, long-term asymptotic behaviors of the SP equation were obtained by Riemann Hilbert method\cite{Boutet2017,Xu2018}. Equation \eqref{a1} has attracted widespread attention in the research of nonlinear theory and has been studied by many scholars. Its mult-soliton solutions, breather solutions and periodic solutions have been studied in \cite{Matsuno2007,Matsuno2008,Saleem2012,Hu2021,Liu2017}.

Parity-time ($\mathcal{PT}$) system maintains invariance under parity transformation and time inversion transformation. The $\mathcal{PT}$-symmetry theory has improved the theory of non-Hermitian Hamiltonians in quantum mechanics and promoted the development of quantum mechanics. In 1998, Bender and Boettcher discovered that a general non-Hermitian Hamiltonian possess real eigenvalues only if it is $\mathcal{PT}$-symmetric\cite{Bender1998}. This theory has been extended to complex domains. Assuming that the potential function {$V(x)$} is complex, which satisfy the condition {$V^*(-x)=V(x)$}, non-Hermitian Hamiltonian {$H=p+V(x)$} is $\mathcal{PT}$-symmetric, where $p$ is a momentum operator. Later, researchers have also extended the concept of symmetry to nonlinear integrable models in mathematical physics. The $\mathcal{PT}$-symmetric integrable system plays an important role in various fields\cite{Ablowitz2013,Ablowitz2017,Priya2019,Stalin2019,Vinayagam2017,Gurses2017,Musslimani2008,Konotop2016,	Zhang2016,Fleury2015,Guo2009}.

The appearance of nonlocal problems is a result of $\mathcal{PT}$-symmetry. Because $\mathcal{PT}$-symmetry leads to the appearance of real spectra for non-Hamilton operators, nonlocal equations remain unchanged under the reverse space-time transformation$(\mathcal{P}: x\rightarrow -x, \mathcal{T}:t\rightarrow -t)$. In recent years, many researchers have studied nonlocal equations. In 2013, Ablowitz and Musslimani proposed the nonlocal nonlinear Schr\"{o}dinger (NNLS) equation\cite{Ablowitz2013}, which was analyzed by the inverse scattering transform method. Then the soliton solutions of the nonlocal NNLS equation also were obtained. After that, several nonlocal equations have been proposed successively, such as the nonlocal mKdV equation, the nonlocal DS equations, the nonlocal sine/sinh-Gordon equation and so on\cite{Ablowitz2017}. The proposal of above equations greatly enriched the research in the field of nonlocal equations. Therefore, studying nonlocal equations has important physical significance, and it is necessary to construct accurate solutions for nonlocal equations.

In order to study the reverse
space-time nonlocal short pulse equation, it is necessary to introduce the two-component short pulse system\cite{Matsuno2011}
\begin{equation}
	\left\{
	\begin{aligned}\label{a2}
	q_{xt}-2q+( qrq_x)_x=0,\\
	r_{xt}-2r+( rqr_x)_x=0,
\end{aligned}
\right.
\end{equation}
where {$q,r$} are complex-value function of variables {$x$} and {$t$}, which satisfy {$r=\sigma q^* (\sigma=\pm 1)$}. $\sigma=1$ represents the case of defocusing, while  $\sigma=-1$ represents the case of focusing. {$q,r$} can be rewrited as {$q(x,t)=a(x,t)+ib(x,t)$}, where {$a(x,t), b(x,t)$} are real functions of variables {$x$} and {$t$}. The solutions of system \eqref{a2} have been studied by many scholars, such as the multi-bright solitons,  multi-breather, and higher-order rogue waves have been gained in \cite{Ling2016}, the multi-dark soliton solution was obtained in \cite{Feng2016}.  

Furtherly, when the nonlocal symmetry condition {$r(x,t)=\sigma q(-x,-t)$} is taken into the system \eqref{a2}, the RST nonlocal short pulse equation has been derived by Yang and Yang\cite{Yang2018}
\begin{equation}\label{a3}
	q_{xt}(x,t)-2q(x,t)+(\sigma q(x,t)q(-x,-t)q_x(x,t))_x=0.
\end{equation}

The infinite conservation laws and the bi-Hamiltonian structure of the equation\eqref{a3} are given in \cite{Brunelli2018}. Multi-soliton solutions of the RST nonlocal short pulse equation have been discussed. The loop soliton solutions of the equation\eqref{a3} were studied by classical Darboux transformation \cite{Hanif2020}. The multi-bright soliton solutions on vanishing background, the multi-dark soliton, multi-breather and higher-order rogue wave solutions corresponding to nonvanishing background were presented through general Darboux transformation \cite{Wang2023}. 

The positon solutions were firstly proposed by Matveev in the Korteweg-de Vries (KdV)  equation in 1992 \cite{Matveev1992}, which are the degenerate cases of the Wronskian formula solution of soliton solutions. The mutual relationship between soliton, positon, and breather has been explained in \cite{Stahlhofen1992}. After that, positon has been discovered in many equations, such as the complex modified KdV (cmKdV) equation\cite{Liu2018}, the sine-Gordon equation\cite{Beutler1993}, the higher-order
Chen-Lee-Liu (HCLL) equation\cite{Hu20212}, the focusing modified Korteweg-de Vries (fmKdV) equation\cite{Xing2017}, the derivative nonlinear Schr\"{o}dinger (dNLS) equation \cite{Song2019} and so on. The positon is a slowly decreasing oscillatory solution  \cite{Song2019,Matveev2002,Hu20212}. Many interesting properties of positon has been introduced, which are different from that of the soliton solutions. For example, the positon possesses super-reflectionless potentials, which is a kind of slowly decaying nonreflective potential\cite{Matveev19922}. The positon is completely transparent in the process of the soliton and positon collison, in other words, the soliton keeps unchanged when colliding with positon, the positon keeps unchanged after colliding with others, but the carrier wave and envelope of positon will derive a ``phase shift"\cite{Dubard2010}. Furthermore, the positon solutions can be obtained through Hirota method and the limit of eigenvalue. Inspired by this significant properties, it is worth to construct the positon solutions of the RST nonlocal short pulse equation.

The positon solutions of the RST nonlocal short pulse equation and the mixed solutions of positon and soliton have not been fully investigated. This is the main aim for this paper. Without loss of generality, in this paper, it only need to consider the focusing case {$(\sigma=-1)$} in equation \eqref{a3}. In this case, it is easy to prove that the potential function {$V(x,t)=-q(x,t)q(-x,-t)$} is obeying the $\mathcal{PT}$-symmetry. Through the transformations of {$x\rightarrow -x,t\rightarrow -t$}, it has {$V(-x,-t)=-q(-x,-t)q(x,t)$}. Obviously, it can be inferred that {$V(x,t)=V(-x,-t)$}, which accords with the $\mathcal{PT}$-symmetry.

This paper is organized as follows. In section 2, the method of constructing classical Darboux Transform will be reviewed. Although the Darboux matrix in this paper is equivalent to that in \cite{Hanif2020}\cite{Wang2023}, the determinant representation of the Darboux matrix is more suitable to take limits on eigenvalues to obtain the positon solutions. Then the $N$-order DT of the RST nonlocal short pulse equation in the $(X,T)-$plane is constructed through $N$-times interation of the 1-fold DT, and the form of determinant of $N$-soliton solution is derived. In section 3, the expression of $N$-soliton solutions for the RST nonlocal short pulse equation in the $(x, t)-$plane is obtained through the hodograph transformation. By taking different eigenvalues, different kinds of the soliton solutions of the RST nonlocal short pulse equation can be generated. In section 4, $N$-order positon solutions can be computed through degenerate Darboux transformation. By performing high-order Taylor expansion of eigenvalues and taking the limit of eigenvalues $(\lambda_{2j-1}\rightarrow \lambda_1,\lambda_{2j}\rightarrow \lambda_2,j=2,...,N)$, degenerate Darboux matrix can be gained. In addition, the mixed solutions of positon and soliton are also studied. Finally, in section 5, some conclusions and discussions are presented.

\section{Darboux transform of the RST nonlocal short pulse equation in the $(X,T)-$plane} \label{1}
\numberwithin{equation}{section}
\subsection{Hodograph transform}
The Lax pair of the RST nonlocal short pulse equation have be given in \cite{Feng2015,Yang2018,Wang2023}
\begin{align}
	\Phi_x=M(x,t;\lambda)\Phi,&&
	\Phi_t=N(x,t;\lambda)\Phi,
\end{align}
with
\begin{align}
	M(x,t;\lambda) =\frac{1}{2\lambda}
	\begin{bmatrix}
		-i & iq_x\\
		-ir_x & i \end{bmatrix},&&
	N(x,t;\lambda) = 
	\begin{bmatrix}
		\frac{ iqr}{2\lambda}+i\lambda & -\frac{i qrq_x}{2\lambda}+q\\
		\frac{irqr_x}{2\lambda}+r& -\frac{ iqr}{2\lambda}-i\lambda  \end{bmatrix},
\end{align}
where $r(x,t)=-q(-x,-t)$ and $\lambda$ is eigenvalue. From the compatibility condition of the RST nonlocal short pulse equation {$\Phi_{xt}=\Phi_{tx}$}, it can be proven that matrices $M$ and $N$ satisfiy the zero curvature equation {$ M_t-N_x+\left[ M,N\right] =0$}.

 Because the equation \eqref{a3} has a WKI type Lax pair, it is necessary to find a hodograph transformation, which maps the independent variables $(x, t)$ to the new variables $(X, T)$. Below equations are referred to the RST nonlocal short pulse equation in the $(X,T)-$plane \cite{Hanif2020}.
\begin{equation}
	\left\{
	\begin{aligned}\label{a4}
		u_{XT}(X,T)=-2u(X,T)\rho(X,T),\\
		\rho_T(X,T)=-(u(X,T)v(X,T))_X,
	\end{aligned}
	\right.
\end{equation}
where $v(X,T)=-u(-X,-T)$. The hodograph transformation between equation\eqref{a3} and equation\eqref{a4} has been defined as follows \cite{Chen2019}
\begin{align}\label{h1}
	dx=\rho dX-uv dT,&&
	dt=-dT,
\end{align}
where $\frac{\partial}{\partial X}=\rho\frac{\partial}{\partial x}, \frac{\partial}{\partial T}=-uv\frac{\partial}{\partial x}-\frac{\partial}{\partial t}, \rho^2-u_Xv_X=1.$
Then the equation\eqref{a3} was equivalently transformed into equation\eqref{a4}. Equation\eqref{a4} has the following AKNS type Lax pair \cite{Wang2023} and satisfies the zero curvature equation $U_T-V_X+[U,V]=0$
\begin{align}\label{a5}
	\Phi_X=U(X,T;\lambda)\Phi,&&
	\Phi_T=V(X,T;\lambda)\Phi,
\end{align}
with
\begin{align}\label{l1}
	U(X,T;\lambda) =\frac{1}{2\lambda}
	\begin{bmatrix}
		-i\rho & -iu_X\\
		iv_X & i\rho \end{bmatrix},&&
	V(X,T;\lambda) = 
	\begin{bmatrix}
		-i\lambda &u\\
		v &i\lambda  \end{bmatrix}.
\end{align}
\subsection{One-fold Darboux transform}
In this section, the classical Darboux transformation \cite{Gu2006,He2006,Li1985} can be constructed  to draw soliton solutions of the RST nonlocal short pulse equation, which is essentially the same as the Darboux transformation used in \cite{Hanif2020}\cite{Wang2023}. However, the form of Darboux matrix $\hat{T}$ for the RST nonlocal short pulse equation is indicated by a determinant representation, which will be more convenient to construct positon solutions. Assume that there is a matrix $\hat{T}$ satisfies $\Phi^{[1]}=\hat{T}\Phi$, and the lax pair $\Phi^{[1]}_X=U^{[1]}\Phi^{[1]}, \Phi^{[1]}_T=V^{[1]}\Phi^{[1]}$, where the forms of $U^{[1]}, V^{[1]}$ are same as that of $U, V$. The compatibility condition $\Phi^{[1]}_{XT}=\Phi^{[1]}_{TX}$ indicates
\begin{equation}
	\left\{
	\begin{aligned}\label{t1}
\hat{T}_X+\hat{T}U=U^{[1]}\hat{T},\\
\hat{T}_T+\hat{T}V=V^{[1]}\hat{T}.
\end{aligned}
\right.
\end{equation}

Similar to the form of DT in \cite{Gu2006,He2006,Li1985}, assume that the one-fold DT has the following form
\begin{equation}
\hat{T}_1=\lambda\begin{bmatrix}
		a_1&b_1\\
		c_1&d_1
	\end{bmatrix}
+\begin{bmatrix}
	a_0&b_0\\
c_0&d_0
\end{bmatrix},
\end{equation}
where $a_1,b_1,c_1,d_1,a_0,b_0,c_0,d_0$ are undetermined functions of $X$ and $T$. At this time, first-order Darboux matrix $T^{[1]}$ is equivalent to one-fold DT $\hat{T}_1$. Substituting $\hat{T}_1$ into equation\eqref{t1} and comparing the coefficients of $\lambda_i (i=0,1,2)$. Obviously, $b_1=c_1=0$, $a_1$ and $b_1$ are constants. Without losing generality, let $a_1=d_1=1$, then it yields
 \begin{align}\label{t4}
 u^{[1]}=u+2ib_0,&&
  v^{[1]}=v-2ic_0,&& \rho^{[1]}=\rho+2ia_{0,X}.
 \end{align}
  Then the expression of the new soliton solutions can be obtained. Furthermore, the one-fold DT $\hat{T}_1$ can be expressed in the form of

\begin{equation}\label{t2}
	\hat{T}_1=\lambda\begin{bmatrix}
		1&0\\
		0&1
	\end{bmatrix}
	+\begin{bmatrix}
		a_0&b_0\\
		c_0&d_0
	\end{bmatrix}.
\end{equation}

Here $a_0,b_0,c_0,d_0$ can be expressed by the eigenfunction $\phi_j,\psi_j (j=1,2)$. $\Phi_1=(\phi_1,\psi_1)^T,\Phi_2=(\phi_2,\psi_2)^T$ are the eigenvectors corresponding to eigenvalues $\lambda_1,\lambda_ 2$ respectively, satisfying $\hat{T}_1(\lambda;\lambda_j)\Phi_j=0,j=1,2$. According to Cramer's law, it yields
\begin{align}\label{t3}
	a_0=-\frac{\left|\begin{matrix}
			\lambda_1\phi_1&\psi_1\\
			\lambda_2\phi_2&\psi_2
	\end{matrix}\right|}{\left| W_1\right|},&&
	b_0=-\frac{\left|\begin{matrix}
		\phi_1&	\lambda_1\phi_1\\
		\phi_2&	\lambda_2\phi_2
		\end{matrix}\right|}{\left| W_1\right|},&&
	c_0=-\frac{\left|\begin{matrix}
			\lambda_1\psi_1&\psi_1\\
			\lambda_2\psi_2&\psi_2
		\end{matrix}\right|}{\left| W_1\right|},&&
	d_0=-\frac{\left|\begin{matrix}
			\phi_1&\lambda_1\psi_1\\
			\phi_2&\lambda_2\psi_2
		\end{matrix}\right|}{\left| W_1\right|},
\end{align}
with
\begin{align}
	W_1=\begin{bmatrix}
	\phi_1&\psi_1\\
	\phi_2&\psi_2
	\end{bmatrix}.
\end{align}
Substituting equation\eqref{t3} into equation\eqref{t2}, the expression of $\hat{T}_1$ can be rewrited as follows
\begin{equation}
	\hat{T}_1=\frac{1}{\left| W_1\right|}\begin{bmatrix}\lambda\left| W_1\right|+\left|\begin{matrix}
			\psi_1&\lambda_1\phi_1\\
			\psi_2&\lambda_2\phi_2
		\end{matrix}\right|&\left|\begin{matrix}
		\lambda_1\phi_1&	\phi_1\\
		\lambda_2\phi_2&	\phi_2
	\end{matrix}\right|\\
\left|\begin{matrix}
	\psi_1&\lambda_1\psi_1\\
	\psi_2&\lambda_2\psi_2
\end{matrix}\right|&\lambda\left| W_1\right|+\left|\begin{matrix}
\lambda_1\psi_1&\phi_1\\
\lambda_2\psi_2&\phi_2
\end{matrix}\right|\end{bmatrix}
		=\frac{1}{\left| W_1\right|}\begin{bmatrix}
		\left|\begin{matrix}
			1&0&\lambda\\
			\phi_1&\psi_1&\lambda_1\phi_1\\
			\phi_2&\psi_2&\lambda_2\phi_2
		\end{matrix}\right|&\left|\begin{matrix}
		0&1&0\\
		\phi_1&\psi_1&\lambda_1\phi_1\\
		\phi_2&\psi_2&\lambda_2\phi_2
	\end{matrix}\right|\\
		\left|\begin{matrix}
			1&0&0\\
			\phi_1&\psi_1&\lambda_1\psi_1\\
			\phi_2&\psi_2&\lambda_2\psi_2
		\end{matrix}\right|&\left|\begin{matrix}
		0&1&\lambda\\
		\phi_1&\psi_1&\lambda_1\psi_1\\
		\phi_2&\psi_2&\lambda_2\psi_2
	\end{matrix}\right|
	\end{bmatrix}.
\end{equation}

Furthertmore, the initial eigenvectors $\Phi_1,\Phi_2$ can be transformed into new forms $\Phi_1^{[1]},\Phi_2^{[1]}$ through Darboux transformation. The determinant representation of $\Phi_1^{[1]},\Phi_2^{[1]}$ as
\begin{equation}\label{phi1}
\Phi_j^{[1]}=\hat{T}_1\Phi_j=\frac{1}{\left| W_1\right|}\begin{bmatrix}
	\left|\begin{matrix}
		\phi_j&\psi_j&\lambda\phi_j\\
		\phi_1&\psi_1&\lambda_1\phi_1\\
		\phi_2&\psi_2&\lambda_2\phi_2
	\end{matrix}\right|\\
	\left|\begin{matrix}
		\phi_j&\psi_j&\lambda\psi_j\\
		\phi_1&\psi_1&\lambda_1\psi_1\\
		\phi_2&\psi_2&\lambda_2\psi_2
	\end{matrix}\right|
\end{bmatrix},j=1,2.
\end{equation}
From formula \eqref{phi1}, it is easy to prove  $\Phi_j^{[1]}=\hat{T}_1\Phi_j=0$, because the elements of the matrix $\Phi_j^{[1]}$ are 0 when $j=1,2$.
Next, considering the reduction condition of the RST nonlocal short pulse equation, two propositions will be derived below

\begin{prop}\label{prop1}
%	\noindent \textbf{Proposition 2.1}:
$\Phi_1=(\phi_1(X,T,\lambda_1),\psi_1(X,T,\lambda_1))^T,\Phi_2=(\phi_2(X,T,\lambda_2),\psi_2(X,T,\lambda_2))^T$ are the eigenvectors of the Lax pair \eqref{l1} corresponding to eigenvalues $\lambda_1,\lambda_ 2$. Here $\phi_1(X,T,\lambda_1)=\psi_1(-X,-T,\lambda_1),\phi_2(X,T,\lambda_2)=-\psi_2(-X,-T,\lambda_2)$, $\lambda_1,\lambda_2$ are arbitrary.
\end{prop}

\begin{proof}
The soliton solutions should satisfy the reduction condition $u(X,T)=-v(-X,-T)$,  $v(X,T)=-u(-X,-T)$. Substituting $\phi_1(X,T,\lambda_1)=\psi_1(-X,-T,\lambda_1),\phi_2(X,T,\lambda_2)=-\psi_2(-X,-T,\lambda_2)$ into $u^{[1]}(X,T)$ in \eqref{t4}
\begin{equation}
	\begin{aligned}
	u^{[1]}(X,T)&=u(X,T)+2i\frac{\phi_1(X,T,\lambda_1)\phi_2(X,T,\lambda_2)(\lambda_1-\lambda_2)}{\psi_2(X,T,\lambda_2)\phi_1(X,T,\lambda_1)-\phi_2(X,T,\lambda_2)\psi_1(X,T,\lambda_1)}\\
	&=-v(-X,-T)-2i\frac{\psi_1(-X,-T,\lambda_1)\psi_2(-X,-T,\lambda_2)(\lambda_1-\lambda_2)}{\phi_1(-X,-T,\lambda_1)\psi_2(-X,-T,\lambda_2)-\psi_1(-X,-T,\lambda_1)\phi_2(-X,-T,\lambda_2)}\\
	&=-v^{[1]}(-X,-T).
\end{aligned}
\end{equation}
Therefore, it can satisfy the reduction condition. Futhermore, $\phi_1(X,T,\lambda_1),\psi_1(X,T,\lambda_1)$ corresponding to eigenvalues $\lambda_1$, while $\phi_2(X,T,\lambda_2),\psi_2(X,T,\lambda_2)$ corresponding to eigenvalues $\lambda_2$, hence, eigenvalues $\lambda_1$ and $\lambda_2$ of nonlocal equation can be arbitrary. 
\end{proof}

\begin{prop}\label{prop2}
There is no reduction condition between the eigenvectors $\Phi_1$ and $\Phi_2$.
\end{prop}
	
\begin{proof}
Substituting eigenvectors $\Phi_1$ and $\Phi_2$ into the Lax pair \eqref{l1}
\begin{align}\label{c4}
	\begin{bmatrix}
		\phi_{1,T}(X,T,\lambda_1)\\
		\psi_{1,T}(X,T,,\lambda_1) \end{bmatrix}=\begin{bmatrix}-i\lambda_1 &u(X,T)\\
		v(X,T) &i\lambda_1 
	\end{bmatrix}\begin{bmatrix}
		\phi_1(X,T,\lambda_1)\\
		\psi_1(X,T,\lambda_1) \end{bmatrix},&&
	\begin{bmatrix}
		\phi_{2,T}(X,T,\lambda_2)\\
		\psi_{2,T}(X,T,\lambda_2) \end{bmatrix}=\begin{bmatrix}-i\lambda_2 &u(X,T)\\
		v(X,T) &i\lambda_2 
	\end{bmatrix}\begin{bmatrix}
		\phi_2(X,T,\lambda_2)\\
		\psi_2(X,T,\lambda_2) \end{bmatrix}.
\end{align}
The above formula can be rewrited into the form of equations, without losing generality, $\Phi_{1,T}=V\Phi_1$ is rewrited as follows 
\begin{equation}
	\left\{
	\begin{aligned}\label{c1}
		\phi_{1,T}(X,T,\lambda_1)&=-i\lambda_1\phi_1(X,T,\lambda_1)+u(X,T)\psi_1(X,T,\lambda_1),\\
		\psi_{1,T}(X,T,\lambda_1)&=v(X,T)\phi_1(X,T,\lambda_1)+i\lambda_1\psi_1(X,T,\lambda_1).
	\end{aligned}
	\right.
\end{equation}
Considering the reduction condition $u(X,T)=-v(-X,-T)$,  By substituting $(X,T)$ for $(-X,-T)$, equations\eqref{c1} become
\begin{equation}
	\left\{
	\begin{aligned}\label{c2}
		\phi_{1,T}(-X,-T,\lambda_1)=-i\lambda_1\phi_1(-X,-T,\lambda_1)-v(X,T)\psi_1(-X,-T,\lambda_1),\\
		\psi_{1,T}(-X,-T,\lambda_1)=-u(X,T)\phi_1(-X,-T,\lambda_1)+i\lambda_1\psi_1(-X,-T,\lambda_1).
	\end{aligned}
	\right.
\end{equation}
Then return equations \eqref{c2} to the form of $\Phi_T=V\Phi$
\begin{equation}\label{c3}
	\begin{bmatrix}
		\psi_{1,T}(-X,-T,\lambda_1)\\
		-\phi_{1,T}(-X,-T,\lambda_1) \end{bmatrix}=\begin{bmatrix}-i(-\lambda_1) &u(X,T)\\
		v(X,T) &i(-\lambda_1) 
	\end{bmatrix}\begin{bmatrix}
		\psi_1(-X,-T,\lambda_1)\\
		-\phi_1(-X,-T,\lambda_1) \end{bmatrix}.
\end{equation}
Comparing formula \eqref{c3} with $\Phi_{2,T}=V\Phi_2$ in \eqref{c4}, it can assume that $\phi_1(X,T,\lambda_1)=-\psi_2(-X,-T,\lambda_2),\psi_1(X,T,\lambda_1)=\phi_2(-X,-T,\lambda_2)$, $\lambda_1=-\lambda_2$. However, if condition $\phi_1(X,T,\lambda_1)=-\psi_2(-X,-T,\lambda_2),\psi_1(X,T,\lambda_1)=\phi_2(-X,-T,\lambda_2)$ is true, the eigenvalue should satisfy $\lambda_1=\lambda_2$, which is inconsistent. If condition $\lambda_1=-\lambda_2$ is true, the eigenfunction should satisfy $\phi_1(X,T,\lambda_1)=-\psi_2(-X,-T,-\lambda_2),\psi_1(X,T,\lambda_1)=\phi_2(-X,-T,-\lambda_2)$, which is also inconsistent. Therefore, there is no reduction condition between the eigenvectors $\Phi_1$ and $\Phi_2$. 
\end{proof}

\subsection{N-fold Darboux transform}
%Similarly, the two-fold DT is obtained by iterating $\hat{T}_1$.
%\begin{equation}
%	\hat{T}_2=\lambda^2\begin{bmatrix}
%		1&0\\
%		0&1
%	\end{bmatrix}
%	+\lambda\begin{bmatrix}
%		a_1^{(2)}&b_1^{(2)}\\
%		c_1^{(2)}&d_1^{(2)}
%	\end{bmatrix}
%+\begin{bmatrix}
%	a_0^{(2)}&b_0^{(2)}\\
%	c_0^{(2)}&d_0^{(2)}
%\end{bmatrix}
%\end{equation}
If the second-order Darboux matrix $\hat{T}^{[2]}$ satisfies $\Phi^{[2]}=\hat{T}^{[2]}\Phi^{[1]}=\hat{T}^{[2]}\hat{T}^{[1]}\Phi$, then two-fold DT $\hat{T}_2$ can be expressed as $\hat{T}^{[2]}\hat{T}^{[1]}$. Two-fold DT is gained through 2-times iterations of one-fold DT. Similarly, the $N$-fold DT of equation\eqref{a4} is obtained by iterating the 1-fold DT N-times, so that $\hat{T}_N=\hat{T}^{[N]}\hat{T}^{[N-1]}...\hat{T}^{[2]}\hat{T}^{[1]}, N\in \mathbb{Z}^+$. The specific form of N-fold Darboux matrix should be
\begin{equation}
	\hat{T}_N=\begin{bmatrix}
		\lambda^{N}+\sum_{i=1}^{N-1}a_i^{(N)}\lambda^{i}&\sum_{i=1}^{N-1}b_i^{(N)}\lambda^{N}\\
		\sum_{i=1}^{N-1}c_i^{(N)}\lambda^{i}&\lambda^{N}+\sum_{i=1}^{N-1}d_i^{(N)}\lambda^{i}
	\end{bmatrix}.
\end{equation}

At the same time, new solutions of the RST nonlocal short pulse equation are
\begin{align}
	u^{[N]}=u+2ib_{N-1}^{(N)},&&
	v^{[N]}=v-2ic_{N-1}^{(N)},&&
	\rho^{[N]}=\rho+2ia_{N-1,X}^{(N)}.
\end{align}

By solving the algebraic equation $\hat{T}_N(\lambda;\lambda_j)\Phi_j=0,j=1,2,...,N$, the determinant form of $a_{N-1}^{(N)},b_{N-1}^{(N)},c_{N-1}^{(N)},d_{N-1}^{(N)}$  can be obtained by Cramer's law 
\begin{align}
	a_{N-1}^{(N)}=\frac{\Delta[N]_{11}}{\left| W_N\right|},&&
b_{N-1}^{(N)}=\frac{\Delta[N]_{12}}{\left| W_N\right|},&&
c_{N-1}^{(N)}=\frac{\Delta[N]_{21}}{\left| W_N\right|},&&
d_{N-1}^{(N)}=\frac{\Delta[N]_{22}}{\left| W_N\right|}.
\end{align}
with
\begin{equation}\nonumber
	\Delta[N]_{11}=-\left|\begin{matrix}
		\phi_1&\psi_1&\lambda_1\phi_1&\lambda_1\psi_1&\cdots&\lambda_1^{N}\phi_1&\lambda_1^{N-1}\psi_1\\
		\phi_2&\psi_2&\lambda_2\phi_2&\lambda_2\psi_2&\cdots&\lambda_2^{N}\phi_2&\lambda_2^{N-1}\psi_2\\
		\phi_3&\psi_3&\lambda_3\phi_3&\lambda_3\psi_3&\cdots&\lambda_3^{N}\phi_3&\lambda_3^{N-1}\psi_3\\
		\phi_4&\psi_4&\lambda_4\phi_4&\lambda_4\psi_4&\cdots&\lambda_4^{N}\phi_4&\lambda_4^{N-1}\psi_4\\
		\vdots&\vdots&\vdots&\vdots&\ddots&\vdots&\vdots\\
		\phi_{2N-1}&\psi_{2N-1}&\lambda_{2N-1}\phi_{2N-1}&\lambda_{2N-1}\psi_{2N-1}&\cdots&\lambda_{2N-1}^{N}\phi_1&\lambda_{2N-1}^{N-1}\psi_{2N-1}\\
		\phi_{2N}&\psi_{2N}&\lambda_{2N}\phi_{2N}&\lambda_{2N}\psi_{2N}&\cdots&\lambda_{2N}^{N}\phi_1&\lambda_{2N}^{N-1}\psi_{2N}
	\end{matrix}\right|,
\end{equation}
\begin{equation}\nonumber
	\Delta[N]_{12}=-\left| \begin{matrix}
		\phi_1&\psi_1&\lambda_1\phi_1&\lambda_1\psi_1&\cdots&\lambda_1^{N-1}\phi_1&\lambda_1^{N}\phi_1\\
		\phi_2&\psi_2&\lambda_2\phi_2&\lambda_2\psi_2&\cdots&\lambda_2^{N-1}\phi_2&\lambda_2^{N}\phi_2\\
		\phi_3&\psi_3&\lambda_3\phi_3&\lambda_3\psi_3&\cdots&\lambda_3^{N-1}\phi_3&\lambda_3^{N}\phi_3\\
		\phi_4&\psi_4&\lambda_4\phi_4&\lambda_4\psi_4&\cdots&\lambda_4^{N-1}\phi_4&\lambda_4^{N}\phi_4\\
		\vdots&\vdots&\vdots&\vdots&\ddots&\vdots&\vdots\\
		\phi_{2N-1}&\psi_{2N-1}&\lambda_{2N-1}\phi_{2N-1}&\lambda_{2N-1}\psi_{2N-1}&\cdots&\lambda_{2N-1}^{N-1}\phi_1&\lambda_{2N-1}^{N}\phi_{2N-1}\\
		\phi_{2N}&\psi_{2N}&\lambda_{2N}\phi_{2N}&\lambda_{2N}\psi_{2N}&\cdots&\lambda_{2N}^{N-1}\phi_1&\lambda_{2N}^{N}\phi_{2N}
	\end{matrix}\right| ,\end{equation}
\begin{equation}\nonumber
	\Delta[N]_{21}=-\left| \begin{matrix}
		\phi_1&\psi_1&\lambda_1\phi_1&\lambda_1\psi_1&\cdots&\lambda_1^{N}\psi_1&\lambda_1^{N-1}\psi_1\\
		\phi_2&\psi_2&\lambda_2\phi_2&\lambda_2\psi_2&\cdots&\lambda_2^{N}\psi_2&\lambda_2^{N-1}\psi_2\\
		\phi_3&\psi_3&\lambda_3\phi_3&\lambda_3\psi_3&\cdots&\lambda_3^{N}\psi_3&\lambda_3^{N-1}\psi_3\\
		\phi_4&\psi_4&\lambda_4\phi_4&\lambda_4\psi_4&\cdots&\lambda_4^{N}\psi_4&\lambda_4^{N-1}\psi_4\\
		\vdots&\vdots&\vdots&\vdots&\ddots&\vdots&\vdots\\
		\phi_{2N-1}&\psi_{2N-1}&\lambda_{2N-1}\phi_{2N-1}&\lambda_{2N-1}\psi_{2N-1}&\cdots&\lambda_{2N-1}^{N}\psi_1&\lambda_{2N-1}^{N-1}\psi_{2N-1}\\
		\phi_{2N}&\psi_{2N}&\lambda_{2N}\phi_{2N}&\lambda_{2N}\psi_{2N}&\cdots&\lambda_{2N}^{N}\psi_1&\lambda_{2N-1}^{N-1}\psi_{2N}
	\end{matrix}\right|, 
\end{equation}
\begin{equation}\nonumber
	\Delta[N]_{22}=-\left|\begin{matrix}
		\phi_1&\psi_1&\lambda_1\phi_1&\lambda_1\psi_1&\cdots&\lambda_1^{N-1}\phi_1&\lambda_1^{N}\psi_1\\
		\phi_2&\psi_2&\lambda_2\phi_2&\lambda_2\psi_2&\cdots&\lambda_2^{N-1}\phi_2&\lambda_2^{N}\psi_2\\
		\phi_3&\psi_3&\lambda_3\phi_3&\lambda_3\psi_3&\cdots&\lambda_3^{N-1}\phi_3&\lambda_3^{N}\psi_3\\
		\phi_4&\psi_4&\lambda_4\phi_4&\lambda_4\psi_4&\cdots&\lambda_4^{N-1}\phi_4&\lambda_4^{N}\psi_4\\
		\vdots&\vdots&\vdots&\vdots&\ddots&\vdots&\vdots\\
		\phi_{2N-1}&\psi_{2N-1}&\lambda_{2N-1}\phi_{2N-1}&\lambda_{2N-1}\psi_{2N-1}&\cdots&\lambda_{2N-1}^{N-1}\phi_1&\lambda_{2N-1}^{N}\psi_{2N-1}\\
		\phi_{2N}&\psi_{2N}&\lambda_{2N}\phi_{2N}&\lambda_{2N}\psi_{2N}&\cdots&\lambda_{2N}^{N-1}\phi_1&\lambda_{2N}^{N}\psi_{2N}
	\end{matrix}\right|,
\end{equation}
\begin{equation}\nonumber
	W_N=\begin{bmatrix}
		\phi_1&\psi_1&\lambda_1\phi_1&\lambda_1\psi_1&\cdots&\lambda_1^{N-1}\phi_1&\lambda_1^{N-1}\psi_1\\
		\phi_2&\psi_2&\lambda_2\phi_2&\lambda_2\psi_2&\cdots&\lambda_2^{N-1}\phi_2&\lambda_2^{N-1}\psi_2\\
		\phi_3&\psi_3&\lambda_3\phi_3&\lambda_3\psi_3&\cdots&\lambda_3^{N-1}\phi_3&\lambda_3^{N-1}\psi_3\\
		\phi_4&\psi_4&\lambda_4\phi_4&\lambda_4\psi_4&\cdots&\lambda_4^{N-1}\phi_4&\lambda_4^{N-1}\psi_4\\
       	\vdots&\vdots&\vdots&\vdots&\ddots&\vdots&\vdots\\
     \phi_{2N}&\psi_{2N}&\lambda_{2N}\phi_{2N}&\lambda_{2N}\psi_{2N}&\cdots&\lambda_{2N}^{N-1}\phi_1&\lambda_{2N}^{N-1}\psi_{2N}
	\end{bmatrix}.
\end{equation}

Similarly as $\hat{T}_1$, the determinant representation of $\hat{T}_N$ can be given as follows
\begin{equation}
	\hat{T}_N=\frac{1}{\left| W_N\right|}\begin{bmatrix}\left|\begin{matrix}
			\eta_1^{[N]}&\lambda^N\\
			W_N&\mu_1^{[N]}
		\end{matrix}\right|&\left|\begin{matrix}
		\eta_2^{[N]}&0\\
		W_N&\mu_1^{[N]}
	\end{matrix}\right|\\
		\left|\begin{matrix}
			\eta_1^{[N]}&0\\
			W_N&\mu_2^{[N]}
		\end{matrix}\right|&\left|\begin{matrix}
		\eta_2^{[N]}&\lambda^N\\
		W_N&\mu_2^{[N]}
	\end{matrix}\right|\end{bmatrix},
\end{equation}
where $\eta_1^{[N]}=\left( 1,0,\lambda,0,\lambda^2,0,...,\lambda^{N-1},0\right),\eta_2^{[N]}=\left( 0,1,0,\lambda,0,\lambda^2,...,0,\lambda^{N-1},\right),\mu_1^{[N]}=\left( \lambda_1^N\phi_1,\lambda_2^N\phi_2,...,\lambda_{2N-1}^N\phi_{2N-1},\lambda_{2N}^N\phi_{2N}\right)^T,\mu_2^{[N]}=\left( \lambda_1^N\psi_1,\lambda_2^N\psi_2,...,\lambda_{2N-1}^N\psi_{2N-1},\lambda_{2N}^N\psi_{2N}\right)^T$.

Then the determinant representation of eigenvector $\Phi_j^{[N]}$ can be obtained from $\Phi_j^{[N]}=\hat{T}_N\Phi_j,j=1,2,..,2N$
\begin{equation}\label{p3}
	\Phi_j^{[N]}=\hat{T}_N\Phi_j=\frac{1}{\left| W_N\right|}\begin{bmatrix}
		\left|\begin{matrix}
			\xi^{[N]}&\lambda^N\phi_j\\
			W_N&\mu_1^{[N]}
		\end{matrix}\right|\\
		\left|\begin{matrix}
			\xi^{[N]}&\lambda^N\psi_j\\
			W_N&\mu_1^{[N]}
		\end{matrix}\right|
	\end{bmatrix},
\end{equation}
where $\xi^{[N]}=\left( \phi_j,\psi_j,\lambda\phi_j,\lambda\psi_j,..,\lambda^{N-1}\phi_j,\lambda^{N-1}\psi_j\right).$
Similarly as $\Phi_j^{[1]}$, it can also be derived that $\Phi_j^{[N]}=\hat{T}_N\Phi_j=0,j=1,2,...2N$.

According to proposition\eqref{prop1} and proposition\eqref{prop2}, it also can gain the following inference

\begin{cor}
 The eigenfunctions satisfy $\phi_{2N-1}(X,T,\lambda_{2N-1})=\psi_{2N-1}(-X,-T,\lambda_{2N-1}),\phi_{2N}(X,T,\lambda_{2N})=-\psi_{2N}(-X,-T,\lambda_{2N})$, and the eigenvalues $\lambda_{2N-1},\lambda_{2N}$ are arbitrary. There is no reduction condition between the eigenvectors $\Phi_{2N-1}$ and $\Psi_{2N}, N\in \mathbb{Z}^+$.
\end{cor}

Further, the new $N$-soliton solutions of the RST nonlocal short pulse equation in the $(X,T)-$plane can be represented by the above determinant:
\begin{equation}
	\left\{
	\begin{aligned}
	u^{[N]}&=u+2i\frac{\Delta[N]_{12}}{\left| W_N\right|},\\
   v^{[N]}&=v-2i\frac{\Delta[N]_{21}}{\left| W_N\right|},\\
	\rho^{[N]}&=\rho+2i\left( \frac{\Delta[N]_{11}}{\left| W_N\right|}\right)_X.
\end{aligned}
\right.
\end{equation}

\section{Soliton solutions of  RST nonlocal short pulse equation in zero background} \label{2}
\numberwithin{equation}{section}
In the previous section, the determinant expression of the $N$-soliton solutions for RST nonlocal short pulse equation in the $(x,t)-$plane have been gained. Although the $N$-soliton solutions of RST nonlocal short pulse equation\eqref{a3} have been disscussed, in order to study the positon solutions, it is necessary to give a determinant representation of the $N$-soliton. According to the hodograph transformation \eqref{h1}, it is easy to yield the expression of soliton solutions for the RST nonlocal short pulse equation in the $(x,t)-$plane
\begin{equation}
\left\{
\begin{aligned}\label{q1}
	 q^{[N]}&=q+2i\frac{\Delta[N]_{12}}{\left| W_N\right|},\\
	r^{[N]}&=r-2i\frac{\Delta[N]_{21}}{\left| W_N\right|},\\
	x^{[N]}&=\int\rho dX-\int uvdT++2i\frac{\Delta[N]_{11}}{\left| W_N\right|}. 
\end{aligned}
\right.
\end{equation}

\subsection{One-soliton solutions in $(x,t)-$plane}
Selecting initial condition  {$u=0$} in equation \eqref{l1}, according to hodograph transformation \eqref{h1}, it yields
\begin{align}
	\rho=\frac{\partial x}{\partial X}=\frac{1}{\sqrt{1+u_Xv_X}}=1,&&
	\frac{\partial x}{\partial T}=uv=0.
\end{align}	
Integrating above equation, it generates $x=X+C$. Without loss of generality, considering $C=0$, it has $x=X$. At that time, $q(x,t)$ and $q(X,T)$ are equivalent solutions. Then the $N$-soliton solution of RST nonlocal short pulse equation \eqref{a3} can be rewrited as 
\begin{equation}
	\left\{
	\begin{aligned}\label{q1}
		q^{[N]}&=q+2i\frac{\Delta[N]_{12}}{\left| W_N\right|},\\
		x^{[N]}&=X+2i\frac{\Delta[N]_{11}}{\left| W_N\right|}. 
	\end{aligned}
	\right.
\end{equation}

When $u=v=0$, the Lax pair \eqref{l1} becomes
\begin{align}\label{f1}
\Phi_X =\frac{1}{2\lambda}
\begin{bmatrix}
	-i & 0\\
	0 & i \end{bmatrix}\Phi,&&
\Phi_T = 
\begin{bmatrix}
	-i\lambda & 0\\
	0& i\lambda  \end{bmatrix}\Phi.
\end{align}

There are 2$N$ eigenvectors {$\Phi_j=(\phi_j,\psi_j)^T (j=1,2,...,2N)$} corresponding to different eigenvalues {$\lambda_1,\lambda_2,...,\lambda_{2N}$}. These independent eigenvectors {$\Phi_j$} are substituted into the above equation \eqref{f1}, then eigenfunctions {$\phi_j,\psi_j$} can be solved,
whose expressions are as follows
\begin{equation}\label{q3}
\begin{bmatrix}
	\phi_j\\
	\psi_j \end{bmatrix}=\begin{bmatrix}A_je^{-\frac{iX}{2\lambda_j}-i\lambda_jT}\\
	B_je^{\frac{iX}{2\lambda_j}+i\lambda_jT}
\end{bmatrix},
\end{equation}
here {$A_j,B_j (j=1,2,...,2N)$} are constants.

When $N=1$, the expressions of one-soliton solutions for the RST nonlocal short pulse equation are following
\begin{equation}
	\left\{
	\begin{aligned}\label{q4}
	q^{[1]}&=\frac{2i\phi_1\phi_2(\lambda_1-\lambda_2)}{\psi_2\phi_1-\psi_1\phi_2},\\
	r^{[1]}&=\frac{2i\psi_1\psi_2(\lambda_1-\lambda_2)}{\psi_2\phi_1-\psi_1\phi_2},\\
	x^{[1]}&=X+\frac{2i(\phi_2\psi_1\lambda_2-\phi_1\psi_2\lambda_1)}{\psi_2\phi_1-\psi_1\phi_2}.
\end{aligned}
\right.
\end{equation}

Considering {$A_1=A_2=B_1=-B_2=1$} in equation\eqref{q3}. It is easy to verify that when {$A_1=A_2=B_1=-B_2=1$}, the soliton solutions satisfies the reduction condition $r(x,t)=-q(-x,-t)$. Then substituting the formula \eqref{q3} into equation \eqref{q4}, the explicit
expression of one-soliton solutions can be expressed as follows:
\begin{equation}
	\left\{
	\begin{aligned}\label{q11}
		q^{[1]}&=-i(\lambda_1-\lambda_2)e^{-\frac{i(\lambda_1+\lambda_2)}{2\lambda_1\lambda_2}X-i(\lambda_1+\lambda_2)T}\sech\left( \frac{i(\lambda_1-\lambda_2)}{2\lambda_1\lambda_2}X-i(\lambda_1-\lambda_2)T\right) ,\\
		r^{[1]}&=i(\lambda_1-\lambda_2)e^{\frac{i(\lambda_1+\lambda_2)}{2\lambda_1\lambda_2}X+i(\lambda_1+\lambda_2)T}\sech\left( \frac{i(\lambda_1-\lambda_2)}{2\lambda_1\lambda_2}X-i(\lambda_1-\lambda_2)T\right) ,\\
		x^{[1]}&=X-i(\lambda_1+\lambda_2)-i(\lambda_1-\lambda_2)\tanh\left( \frac{i(\lambda_1-\lambda_2)}{2\lambda_1\lambda_2}X-i(\lambda_1-\lambda_2)T\right) .
	\end{aligned}
	\right.
\end{equation}

\begin{figure}[!htbp]
	\centering
	\subfigure[]{\includegraphics[height=4.5cm,width=4.5cm]{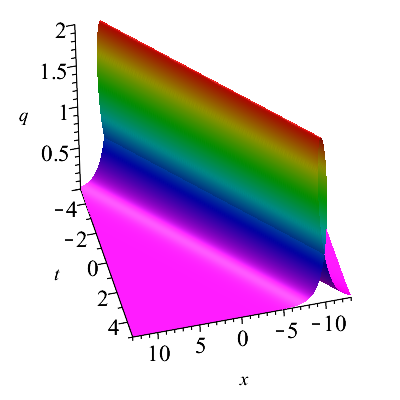}}
	\subfigure[]{\includegraphics[height=4.5cm,width=4.5cm]{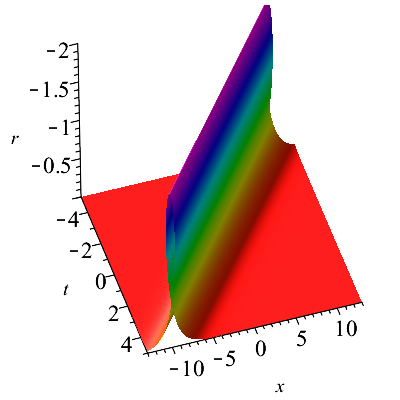}}\\
	\caption{3D plots of the bounded one-soliton for the RST nonlocal short pulse equation in $(x,t)-$plane with parameters $\alpha_1=1,\alpha_2=-1$ from case 3.1. Panel (a) is the plot of $q^{[1]}$; panel (b) is the plot of $r^{[1]}$.}
	\label{fig1s1}
\end{figure}

Because equation\eqref{a3} is nonlocal, the values of {$\lambda_1,\lambda_2$} are  arbitrary. We assume that $\lambda_1=i\alpha_1, \lambda_2=i\alpha_2$, where $\alpha_1,\alpha_2$ are positive real constants and $\alpha_1\not=\alpha_2$. Different solutions can be obtained when $\alpha_1,\alpha_2$ are different. Following cases can be classified:

\noindent \textbf{Case 3.1}: When $\alpha_1=-\alpha_2$, eigenvalue $\lambda_1$ and $\lambda_2$ are conjugate in this case. One has the one-loop soliton solutions in $(x,t)-$plane. The shape of the one-loop soliton in $(x,t)-$plane is represented in the Fig.\ref{fig1s1} through 3D plot, (a) is the shape of $q^{[1]}$ and (b) is the shape of $r^{[1]}$. The parameters are taken as Fig.\ref{fig1s1} with $\alpha_1=1,\alpha_2=-1$. In this time, one-soliton is bounded, the one-soliton solutions given by equation \eqref{a3} are same as those given by equation\eqref{a2}. In other words, the soliton shown in the Fig.\ref{fig1s1} is both the solution of equation\eqref{a3} and that of equation\eqref{a2}. 

\begin{figure}[!htbp]
		\centering
		\subfigure[]{\includegraphics[height=4.5cm,width=4.5cm]{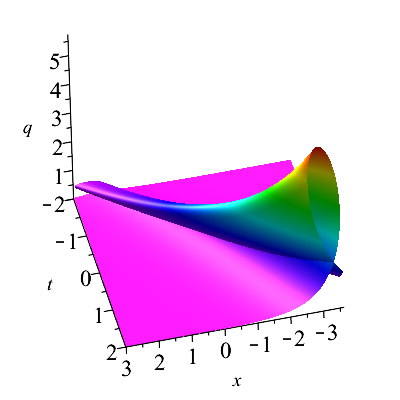}}
		\subfigure[]{\includegraphics[height=4.5cm,width=4.5cm]{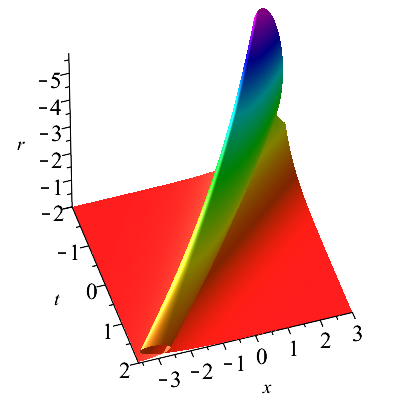}}
		\subfigure[]{\includegraphics[height=4.5cm,width=4.5cm]{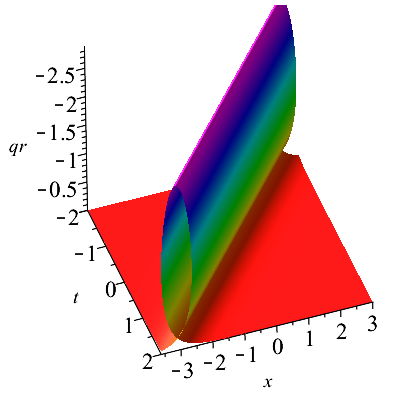}}\\
		\caption{3D plots of the unbounded one-soliton for the RST nonlocal short pulse equation in $(x,t)-$plane with parameters $\alpha_1=0.7,\alpha_2=-1$ from case 3.2. Panel (a) is the plot of $q^{[1]}$; panel (b) is the plot of $r^{[1]}$; panel (c) is the plot of $q^{[1]}r^{[1]}$.}
		\label{fig1s2}
\end{figure}

\noindent \textbf{Case 3.2}: When $\left| \alpha_1\right| \not =\left| \alpha_2\right| $, one generates the unbounded one-loop soliton solution in $(x,t)-$plane, whose shape is plotted in Fig.\ref{fig1s2}, (a) is the shape of $q^{[1]}$, (b) is the shape of $r^{[1]}$ and (c) is the shape of $q^{[1]}r^{[1]}$. The parameters are choosen as Fig.\ref{fig1s2} with $\alpha_1=0.7,\alpha_2=-1$. It is easy to find that $q^{[1]}, r^{[1]}$ are unbounded soliton. $q^{[1]}\rightarrow +\infty$ but $r^{[1]}\rightarrow 0$ when $t\rightarrow +\infty$. However, when $t\rightarrow -\infty$, $q^{[1]}\rightarrow 0$ but $r^{[1]}\rightarrow -\infty$.  $\left| q^{[1]}\right|$ loss along the negative $t$-direction but grow along the nonnegative $t$-direction, $\left| r^{[1]}\right|$ is opposite. Assume that the potential function $V^{[1]}(x,t)=q^{[1]}(x,t)r^{[1]}(x,t)$, $V^{[1]}$ is bounded, which can be seen from Fig.\ref{fig1s2}(c), because grow and decrease are balanced. 
	
\subsection{Two-soliton solutions in $(x,t)-$plane} 

When $N=2$ in equation\eqref{q1}, the expression of two-soliton solutions in zero background are as follows:
\begin{flalign}
q^{[2]}&=-\frac{2i}{ \left| W_2\right|}\left| \begin{matrix}
	\phi_1&\psi_1&\lambda_1\phi_1&\lambda_1^2\phi_1\\
	\phi_2&\psi_2&\lambda_2\phi_2&\lambda_2^2\phi_2\\
	\phi_3&\psi_3&\lambda_3\phi_3&\lambda_3^2\phi_3\\
	\phi_4&\psi_4&\lambda_4\phi_4&\lambda_4^2\phi_4
\end{matrix}\right|,\\
r^{[2]}=&\frac{2i}{ \left| W_2\right|}\left| \begin{matrix}
	\phi_1&\psi_1&\lambda_1^2\psi_1&\lambda_1\psi_1\\
	\phi_2&\psi_2&\lambda_2^2\psi_2&\lambda_2\psi_2\\
	\phi_3&\psi_3&\lambda_3^2\psi_3&\lambda_3\psi_3\\
	\phi_4&\psi_4&\lambda_4^2\psi_4&\lambda_4\psi_4
\end{matrix}\right|,\\
\end{flalign}
\begin{equation}
x^{[2]}=X+\frac{-2i}{ \left| W_2\right|}\left| \begin{matrix}
	\phi_1&\psi_1&\lambda_1^2\phi_1&\lambda_1\psi_1\\
	\phi_2&\psi_2&\lambda_2^2\phi_2&\lambda_2\psi_2\\
	\phi_3&\psi_3&\lambda_3^2\phi_3&\lambda_3\psi_3\\
	\phi_4&\psi_4&\lambda_4^2\phi_4&\lambda_4\psi_4
\end{matrix}\right|,
\end{equation}
with
\begin{flalign}
	W_2&=\begin{bmatrix}
		\phi_1&\psi_1&\lambda_1\phi_1&\lambda_1\psi_1\\
		\phi_2&\psi_2&\lambda_2\phi_2&\lambda_2\psi_2\\
		\phi_3&\psi_3&\lambda_3\phi_3&\lambda_3\psi_3\\
		\phi_4&\psi_4&\lambda_4\phi_4&\lambda_4\psi_4
	\end{bmatrix}.
\end{flalign}

\begin{figure}[!htbp]
	\centering
	\subfigure[]{\includegraphics[height=4.5cm,width=4.5cm]{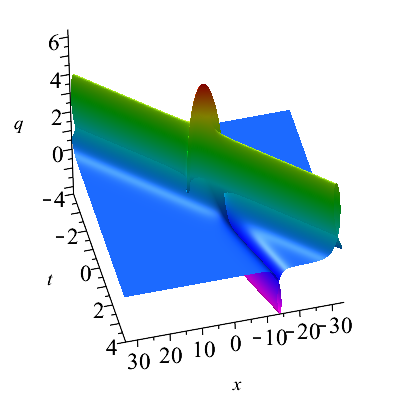}}
	\subfigure[]{\includegraphics[height=4.5cm,width=4.5cm]{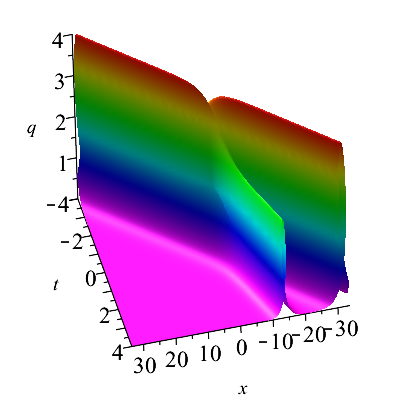}}\\
	\caption{3D plots of the bounded two-soliton for the RST nonlocal short pulse equation in $(x,t)-$plane from case 3.3. Panel (a) is the plot of $q^{[2]}$ with parameters $\alpha_1=1.2,\alpha_2=-1.2,\alpha_3=2,\alpha_4=-2$; panel (b) is the plot of $q^{[2]}$ with parameters $\alpha_1=-1.2,\alpha_2=1.2,\alpha_3=2,\alpha_4=-2$.}
	\label{fig2s1}
\end{figure}

Similar to the cases of one-soliton, it is necessary to consider different cases of two-soliton. When different eigenvalues are taken, it yields the following different cases:

\noindent \textbf{Case 3.3}: When $\alpha_1=-\alpha_2, \alpha_3=-\alpha_4$, the bounded two-loop soliton solution can be obtained as shown in Fig\ref{fig2s1}. If $\alpha_1$ and $\alpha_3$ are both nonnegative, the parameters selected in Fig\ref{fig2s1} (a) are $\alpha_1=1.2,\alpha_2=-1.2,\alpha_3=2,\alpha_4=-2$. In this case, one wave of two-soliton is negative while another wave is nonegative. If one of $\alpha_1, \alpha_3$ is negative but another is nonnegative, the parameters choosen in Fig\ref{fig2s1} (b) are $\alpha_1=1.2,\alpha_2=-1.2,\alpha_3=-2,\alpha_4=2$. Differing from the two-soliton shown in Fig\ref{fig2s1} (a),  two waves of two-soliton are both nonegative in Fig\ref{fig2s1} (b). Focus on Fig.\ref{fig2s1}, during the transmission process, two loop solitons undergo an elastic collision, where the two waves interact with each other and then separate from each other. The velocity and shape of the two waves do not change after the collision, so this collision is considered an elastic collision.

\begin{figure}[!htbp]
	\centering
	\subfigure[]{\includegraphics[height=4.5cm,width=4.5cm]{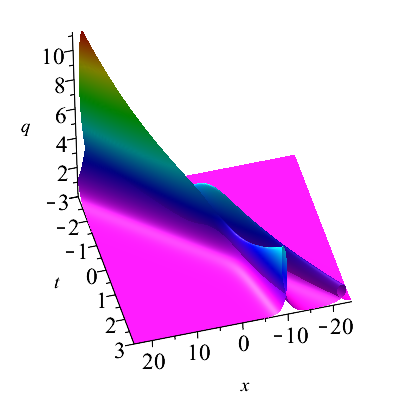}}
	\subfigure[]{\includegraphics[height=4.5cm,width=4.5cm]{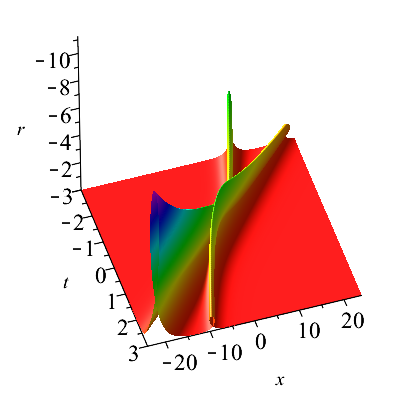}}
	\subfigure[]{\includegraphics[height=4.5cm,width=4.5cm]{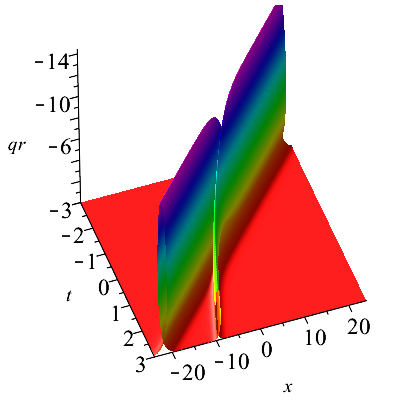}}\\
	\caption{3D plots of the unbounded two-soliton for the RST nonlocal short pulse equation in $(x,t)-$plane with parameters $\alpha_1=-1.2,\alpha_2=1,\alpha_3=2,\alpha_4=-1.8$ from case 3.4. Panel (a) is the plot of $q^{[2]}$ ; panel (b) is the plot of $r^{[2]}$; panel (c) is the plot of $q^{[2]}r^{[2]}$.}
	\label{fig2s2}
	\subfigure[]{\includegraphics[height=4.5cm,width=4.5cm]{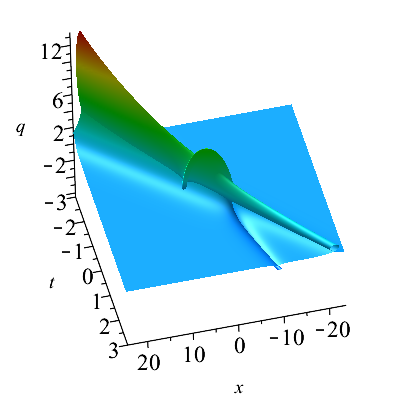}}
	\subfigure[]{\includegraphics[height=4.5cm,width=4.5cm]{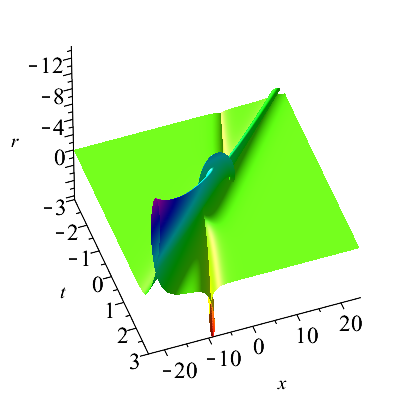}}
	\subfigure[]{\includegraphics[height=4.5cm,width=4.5cm]{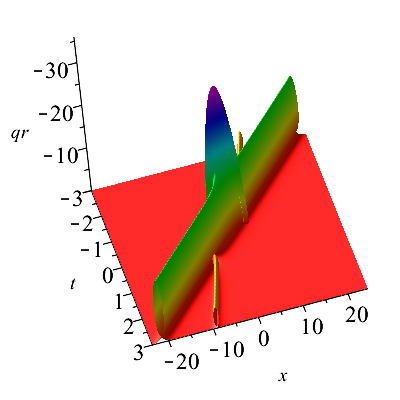}}\\
	\caption{3D plots of the unbounded two-soliton for the RST nonlocal short pulse equation in $(x,t)-$plane with parameters $\alpha_1=1.2,\alpha_2=-1,\alpha_3=2,\alpha_4=-1.8$ from case 3.4. Panel (a) is the plot of $q^{[2]}$ ; panel (b) is the plot of $r^{[2]}$; panel (c) is the plot of $q^{[2]}r^{[2]}$.}
	\label{fig2s3}
\end{figure}

\noindent \textbf{Case 3.4}: When $\left| \alpha_1\right| \not =\left| \alpha_2\right|, \left| \alpha_3\right| \not =\left| \alpha_4\right| $, unbounded two-loop soliton solution of the RST nonlocal short pulse equation \eqref{a3} can be obtained. Similar with case 3.3, if $\alpha_1$ and $\alpha_3$ are both nonnegative, an unbounded two-soliton shown in Fig.\ref{fig2s2} with the parameters $\alpha_1=-1.2,\alpha_2=1,\alpha_3=2,\alpha_4=-1.8$, (a) is the shape of $q^{[2]}$, (b) is the shape of $r^{[2]}$, (c) is the shape of $q^{[2]}r^{[2]}$. Fig.\ref{fig2s2} shows two waves of $q^{[2]}$ are both nonegative, one wave growing when $t$ becomes negative while another is opposite. Futhermore, if one of $\alpha_1, \alpha_3$ is negative but another is nonnegative, the parameters in Fig.\ref{fig2s3} are $\alpha_1=-1.2,\alpha_2=1,\alpha_3=-2,\alpha_4=1.8$. Compared to Fig.\ref{fig2s2}, it can be seen from Fig.\ref{fig2s3}, one wave of $q^{[2]}$ is nonegative but another is negative. The absolute values of two waves are both growing when $t$ becomes negative. It is obvious that $q^{[2]}$ and $r^{[2]}$ are unbounded soliton, but $q^{[2]}r^{[2]}$ become bounded.

Furthermore, higher-order solitons {$q^{[N]}$} can also be calculated by the similar method, which will be more complicated.

\section{The positon solutions and mixed solution of the RST nonlocal short pulse equation} \label{3}
\numberwithin{equation}{section}

\subsection{Positons of the RST nonlocal short pulse equation}
Positon solutions is derived from the degenerate DT and the higher-order Taylor expansion. Because equation\eqref{a3} is nonlocal, the degenerate $N$-fold Darboux matrix of the nonlocal equation can be gained by setting eigenvalues $\lambda_{2j-1}= \lambda_1+\epsilon,\lambda_{2j}=\lambda_2+\epsilon (j=2,...,N)$, where {$\epsilon$} is a small complex parameter. Because $\lambda_1,\lambda_2$ are arbitrary parameters, there exist no reduction conditions between $\lambda_1,\lambda_2$, the degenerate Darboux matrix of nonlocal equation is different from that of local equation. Obviously, choosing $N=2$ in equation\eqref{q1}, the denominator of two-soliton solution is 0 when {$\lambda_3 = \lambda_1, \lambda_4 = \lambda_2$}, then the two-soliton solution will be the {$\frac{0}{0}$}-infinitive. Similarly, for larger $N$, $N$-degenerate solution can also be obtained by using the similar method, which will be more complex. It should be noting that positon solutions of the RST nonlocal short pulse equation can be gained from a zero ``seed" solution by substituting $\Phi_j(j=1,2,...,N)$ into the degenerate Darboux matrix and the higher-order Taylor expansion of eigenvalues $\lambda_{2j-1}= \lambda_1+\epsilon,\lambda_{2j}=\lambda_2+\epsilon (j=2,...,N)$.

\begin{prop}
 Let $\lambda_{2j-1}= \lambda_1+\epsilon,\lambda_{2j}=\lambda_2+\epsilon,j=2,...,N$, the $N$-positon solution $q^{[N]}_p,r^{[N]}_p$ of the RST nonlocal short pulse equation based on zero ``seed" solution $q=0$ can be expressed as follows
\begin{equation}
	\left\{
	\begin{aligned}\label{p1}
		q^{[N]}_{p}&=\frac{\Delta[N]_{12}^{'}}{\left|W_N^{'}\right|},\\
		r^{[N]}_{p}&=\frac{\Delta[N]_{21}^{'}}{\left|W_N^{'}\right|},\\
		x^{[N]}_{p}&=X+\frac{\Delta[N]_{11}^{'}}{\left|W_N^{'}\right|},
\end{aligned}
\right.
\end{equation}
where
\begin{gather}\nonumber
		\Delta[N]_{11}^{'}=\left( \frac{\partial^{N_i}}{\partial\epsilon^{N_i}}|_{\epsilon=0} \left( \Delta[N]_{11}\right) _{ij}\left( \lambda_k+\epsilon\right) \right) _{2N\cdot 2N},\\
		\nonumber
		\Delta[N]_{12}^{'}=\left( \frac{\partial^{N_i}}{\partial\epsilon^{N_i}}|_{\epsilon=0} \left( \Delta[N]_{12}\right) _{ij}\left( \lambda_k+\epsilon\right) \right) _{2N\cdot 2N},\\
		\nonumber
		\Delta[N]_{21}^{'}=\left( \frac{\partial^{N_i}}{\partial\epsilon^{N_i}}|_{\epsilon=0} \left( \Delta[N]_{21}\right) _{ij}\left( \lambda_k+\epsilon\right) \right) _{2N\cdot 2N},\\
		\nonumber
		\left( W_N^{'}\right) _{ij}=\left( \frac{\partial^{N_i}}{\partial\epsilon^{N_i}}|_{\epsilon=0} \left( W_N\right) _{ij}\left( \lambda_k+\epsilon\right) \right) _{2N\cdot 2N},
	\end{gather}
	here $N_i=\left[ \frac{i-1}{2}\right] $, $[i]$ expresses the floor function of $i$. When $i$ is odd, $k=1$, and when $i$ is even, $k=2$.
\end{prop}

	Specifically, when $N=1$ in Proposition 1, one-positon can be obtained, which is equal to one-soliton solutions\eqref{q11}. Therefore, the two-positon solution is the first non-trivial positon solution. When $N=2$, substituting $\lambda_1=i\alpha_1, \lambda_2=i\alpha_2$, the explicit expression of 2-positon can be given as follows
\begin{equation}
	\left\{
	\begin{aligned}\label{p2}
	q^{[2]}_{p}&=\frac{A_1+A_2}{2\left( 1+i \right) {\alpha_1}^{2}{\alpha_1}^{2}\cosh(H_1-H_2)+A_4},\\
	x^{[2]}_{p}&=x+\frac{2(1+i)\alpha_1^2\alpha_2^2\left[  \left( 2\alpha_1+2\alpha_1+X \right)  \cosh(H_1-H_2)+\left( 2\alpha_1-2\alpha_2\right) \sinh(H_1-H_2)\right] +A_3}{2\left( 1+i \right) {\alpha_1}^{2}{\alpha_1}^{2}\cosh(H_1-H_2)+A_4},
\end{aligned}
\right.
\end{equation}
with
\begin{flalign}
	\begin{split}
	A_1&=a_1e^{H_1},
	A_2=a_2e^{H_2},
	A_3=w_1+w_2+w_3+w_4+w_5,
	A_4=2k_1+k_2,\\
a_1&=\left( -4-4\,i \right)  \left(  \left( T{\alpha_2}^{2}+\frac{X}{2} \right) 
\alpha_1- \left( T{\alpha_2}^{2}+\frac{X}{2}+\alpha_2 \right) \alpha_2 \right) 
\left( \alpha_1-\alpha_2 \right) {\alpha_1}^{2}
, \\
a_2&=\left( 4+4\,i \right) {\alpha_2}^{2} \left( T{\alpha_1}^{3}+ \left( -T
\alpha_2+1 \right) {\alpha_1}^{2}+\frac{1}{2}\,X\alpha_1-\frac{1}{2}\,X\alpha_2 \right) 
\left( \alpha_1-\alpha_2 \right)
, \\
	w_1&= \left( 2+2\,i \right) {\alpha_2}^{2}X \left( 2\,T{\alpha_2}^{3}+ \left( T
	X+1 \right) {\alpha_2}^{2}+X\alpha_2+\frac{1}{2}\,{X}^{2} \right) 
	, \\
	w_2&= \left(  \left( -8-8\,i \right) {T}^{2}{\alpha_2}^{3}+ \left( 4+4\,i
	\right) T \left( TX+2 \right) {\alpha_2}^{2}- \left( 4+4\,i \right) XT
	\alpha_2+ \left( 2+2\,i \right)  \left( TX+1 \right) X \right) {\alpha_1}^
	{4}
	,\\
	w_3&= \left(  \left( -8-8\,i \right) {T}^{2}{\alpha_2}^{4}- \left( 8+8\,i
	\right) T \left( TX+2 \right) {\alpha_2}^{3}+ \left( 4+4\,i \right) {
		\alpha_2}^{2}- \left( 4+4\,i \right)  \left( TX+1 \right) X\alpha_2+
	\left( 2+2\,i \right) {X}^{2} \right) {\alpha_1}^{3}
	,\\
	w_4&=\left(  \left( 8+8\,i \right) {T}^{2}{\alpha_2}^{5}+ \left( 4+4\,i
	\right) T \left( TX+2 \right) {\alpha_2}^{4}+ \left( 4+4\,i \right) {
		\alpha_2}^{3}+ \left( 4+4\,i \right) X \left( TX+\frac{3}{2} \right) {\alpha_2}^{2
	}- \left( 2+2\,i \right) {X}^{2}\alpha_2+ \left( 1+i \right) {X}^{3}
	\right) {\alpha_1}^{2}
	,\\
	w_5&=\left( 8+8\,i \right) T \left( T{\alpha_2}^{2}+\frac{X}{2} \right) {\alpha_1}^{5}
	- \left( 4+4\,i \right) \alpha_2\,X \left( T{\alpha_2}^{3}+ \left( TX+1
	\right) {\alpha_2}^{2}+\frac{1}{2}\,X\alpha_2+\frac{1}{2}\,{X}^{2} \right) \alpha_1
	,\\
	k_1&=\left( 2+2\,i \right) {\alpha_2}^{2} \left( \alpha_1-\alpha_2 \right) ^{2}{
		\alpha_1}^{2}{T}^{2}+ \left( 1+i \right) X \left( {\alpha_1}^{2}+{\alpha_2}^
	{2} \right)  \left( \alpha_1-\alpha_2 \right) ^{2}T
	,\\
		k_2&= \left(  \left( 2+2\,i \right) {\alpha_2}^{2}+ \left( 1+i \right) {X}^{2
		} \right) {\alpha_1}^{2}- \left( 2+2\,i \right) \alpha_2\,{X}^{2}\alpha_1+
		\left( 1+i \right) {\alpha_2}^{2}{X}^{2}
		,\\
	H_1&={\frac {2\,T{\alpha_1}^{2}-X}{\alpha_1}},
		H_2={\frac {2T{\alpha_2}^{2}-X}{\alpha_2}}.
\end{split}&
\end{flalign}
Differing from soliton solutions, the expression of the position solution is composed by blend of exponential functions and polynomials about $X$ and $T$, so that they also have different properties. 
Next, the properties of the 2-positon solutions will be discussed when different eigenvalues are taken.

\begin{figure}[!htbp]
	\centering
	\subfigure[]{\includegraphics[height=4.5cm,width=4.5cm]{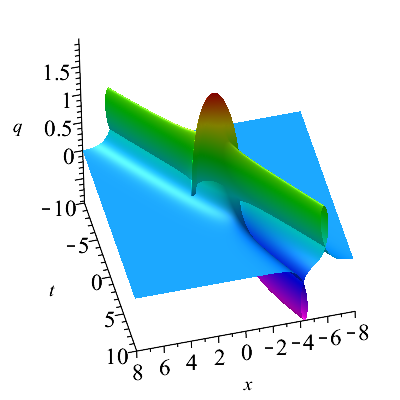}}
	\subfigure[]{\includegraphics[height=4.5cm,width=4.5cm]{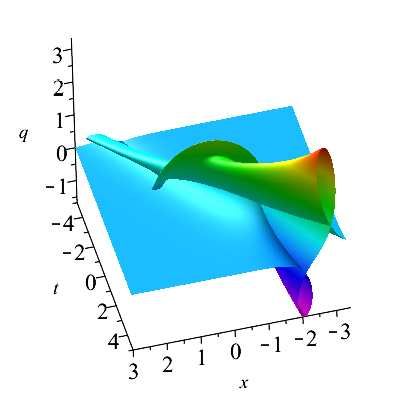}}
    \subfigure[]{\includegraphics[height=4.5cm,width=4.5cm]{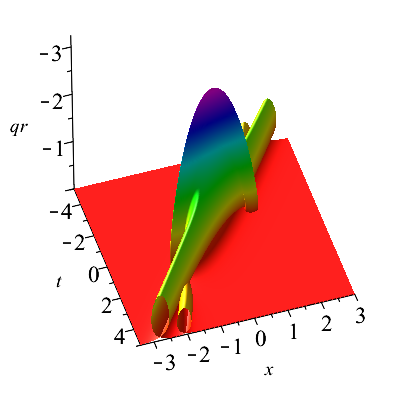}}	\\
	\caption{3D plots of two-positon solutions for the RST nonlocal short pulse equation in $(x,t)-$plane from case 4.1. Panel (a) is a bounded two-positon $q_p^{[2]}$ with parameters $\alpha_1=0.5$ and $\alpha_2=-0.5$; panel (b) is an unbounded two-positon $q_p^{[2]}$ with parameters $\alpha_1=0.4$ and $\alpha_2=-0.5$; panel (c) is the plot of $q_p^{[2]}r_p^{[2]}$ with parameters $\alpha_1=0.4$ and $\alpha_2=-0.5$.}
	\label{fig2p1}
	\subfigure[]{\includegraphics[height=4.5cm,width=4.5cm]{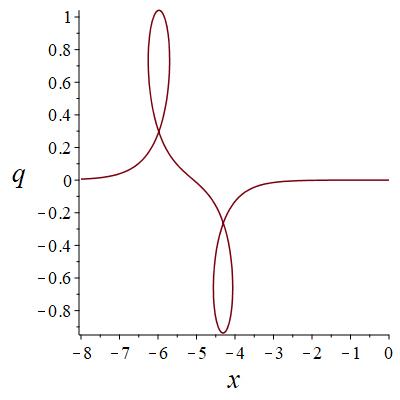}}
	\subfigure[]{\includegraphics[height=4.5cm,width=4.5cm]{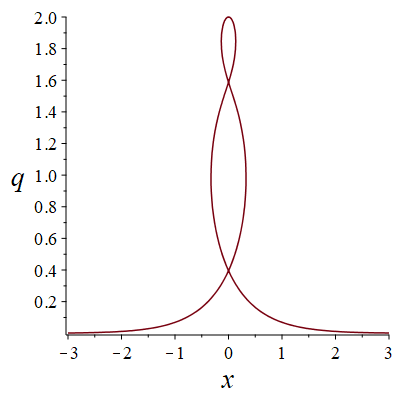}}
	\subfigure[]{\includegraphics[height=4.5cm,width=4.5cm]{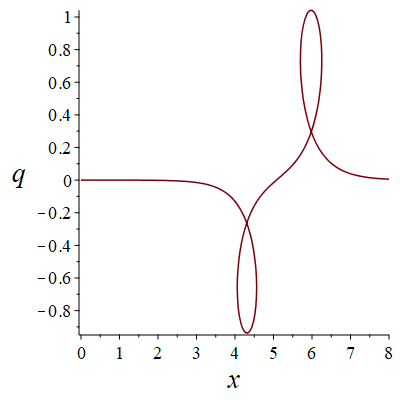}}
	\caption{Profiles of bounded two-positon solution $q_p^{[2]}$ for the RST nonlocal short pulse equation in $(x,t)-$plane with parameters $\alpha_1=0.5, \alpha_2=-0.5$. Panel (a) is the profile at $t=10$; panel (b) is the profile at $t=0$; panel (c) is the profile at $t=-10$.}
	\label{fig2p2}
	\subfigure[]{\includegraphics[height=4.5cm,width=4.5cm]{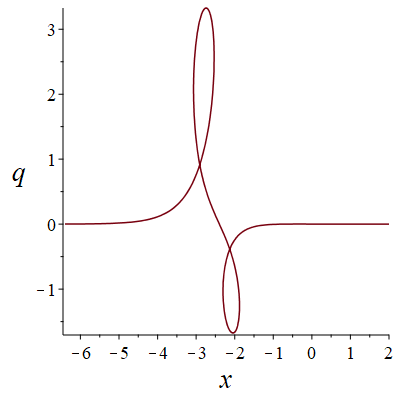}}
	\subfigure[]{\includegraphics[height=4.5cm,width=4.5cm]{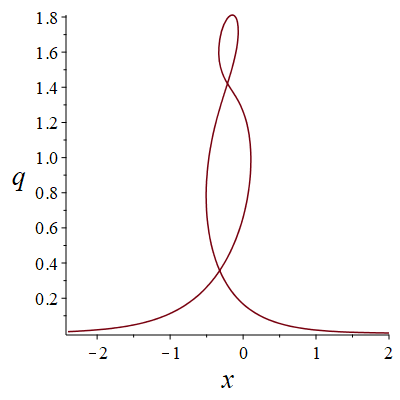}}
	\subfigure[]{\includegraphics[height=4.5cm,width=4.5cm]{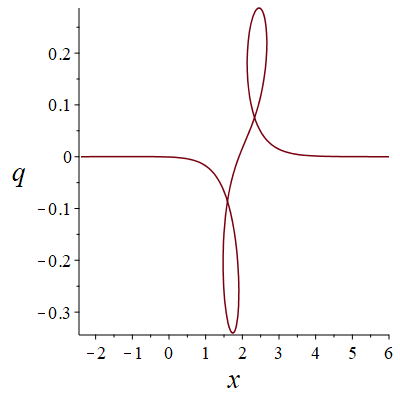}}
	\caption{Profiles of unbounded two-positon solution $q_p^{[2]}$ for the RST nonlocal short pulse equation in $(x,t)-$plane with parameters $\alpha_1=0.4, \alpha_2=-0.5$. Panel (a) is the profile at $t=5$; panel (b) is the profile at $t=0$; panel (c) is the profile at $t=-5$.}
	\label{fig2p3}
\end{figure}

\noindent \textbf{Case 4.1}: When $N=2$, the bounded two-positon solution of the RST nonlocal short pulse equation \eqref{a3} can be obtained with $\alpha_2=-\alpha_1$. At that time, $\lambda_1=\lambda_2^*$, the two-positon is also the solution to equation\eqref{a2}. The Fig.\ref{fig2p1} (a) shows some characteristics of the two-positon for the equation \eqref{a3} in $(x,t)-$plane. Fig.\ref{fig2p2} shows the profile of the bounded two-positon at (a) $t=10$ (b) $t=0$ and (c) $t=-10$ respectively. The parameters are choosen with $\alpha_1=0.5, \alpha_2=-0.5$. It is obvious that loop positon is generated in $(x,t)-$plane. By analyzing legends, it can be seen that the two-positon is a set of slowly varying curves, which is different from two-soliton. Futhermore, before and after the collision, the shape and amplitude of the two-positon keeps unchanged, but there exist a ``phase shift" during the collision. Moreover, the unbounded two-positon solution of the RST nonlocal short pulse equation \eqref{a3} can be generated when  $\left| \alpha_1\right| \not =\left| \alpha_2\right| $. The parameters are choosen as Fig.\ref{fig2p1} (b) with $\alpha_1=0.4, \alpha_2=-0.5$. the profile of the unbounded two-positon is shown in Fig.\ref{fig2p3} at (a) $t=5$ (b) $t=0$ and (c) $t=-5$. Two waves both approach 0 along the negative $t$-direction. Similar with the soliton, $V_p^{[2]}=q_p^{[2]}r_p^{[2]}$ is bounded which makes unbounded two-positon $q_p^{[2]}$ and$ r_p^{[2]}$ balanced.

\begin{figure}[!htbp]
	\centering
	\subfigure[]{\includegraphics[height=4.5cm,width=4.5cm]{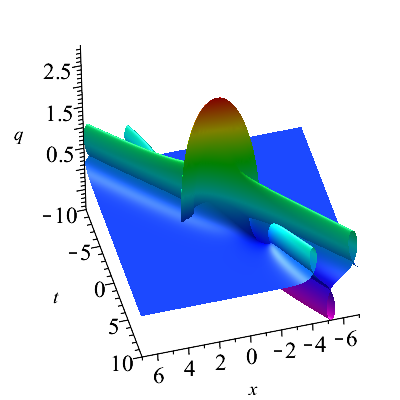}}
	\subfigure[]{\includegraphics[height=4.5cm,width=4.5cm]{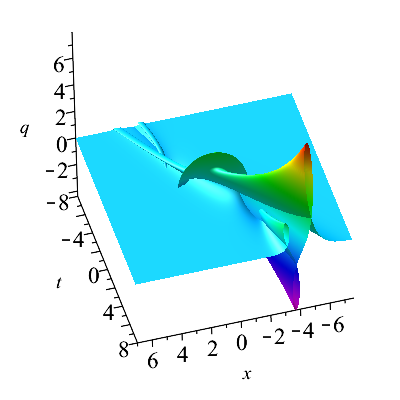}}
	\subfigure[]{\includegraphics[height=4.5cm,width=4.5cm]{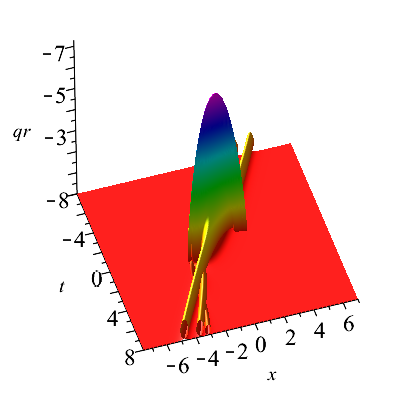}}
	\caption{3D plots of three-positon solutions $q_p^{[3]}$ for the RST nonlocal short pulse equation in $(x,t)-$plane from case 4.2. Panel (a) is a bounded three-positon with parameters $\alpha_1=0.5$ and $\alpha_2=-0.5$; panel (b) is an unbounded three-positon with parameters $\alpha_1=0.4$ and $\alpha_2=-0.5$; panel (c) is the plot of $q_p^{[3]}r_p^{[3]}$ with parameters $\alpha_1=0.4$ and $\alpha_2=-0.5$.}
	\label{fig3p1}
	\subfigure[]{\includegraphics[height=4.5cm,width=4.5cm]{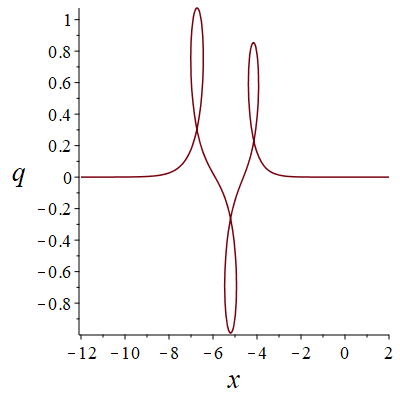}}
	\subfigure[]{\includegraphics[height=4.5cm,width=4.5cm]{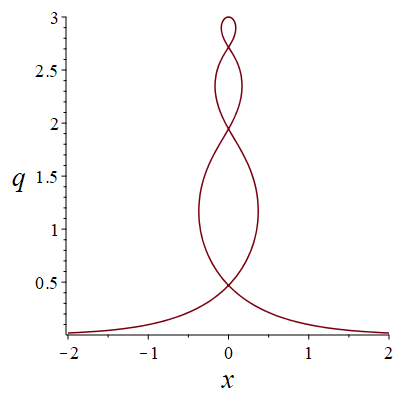}}
	\subfigure[]{\includegraphics[height=4.5cm,width=4.5cm]{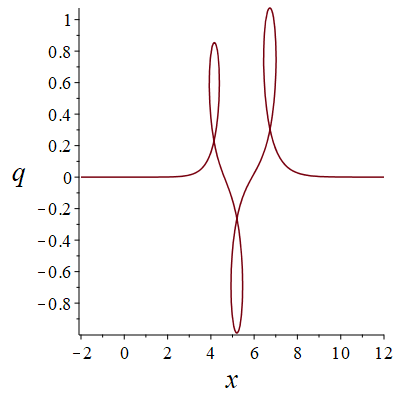}}\\
	\caption{Profiles of bounded three-positon solution $q_p^{[3]}$ for the RST nonlocal short pulse equation in $(x,t)-$plane with parameters $\alpha_1=0.5, \alpha_2=-0.5$. Panel (a) is the profile at $t=10$; panel (b) is the profile at $t=0$; panel (c) is the profile at $t=-10$.}
	\label{fig3p2}
	\subfigure[]{\includegraphics[height=4.5cm,width=4.5cm]{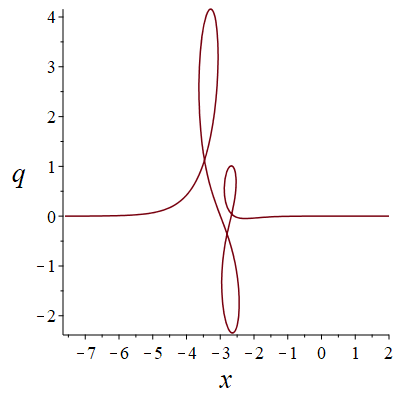}}
	\subfigure[]{\includegraphics[height=4.5cm,width=4.5cm]{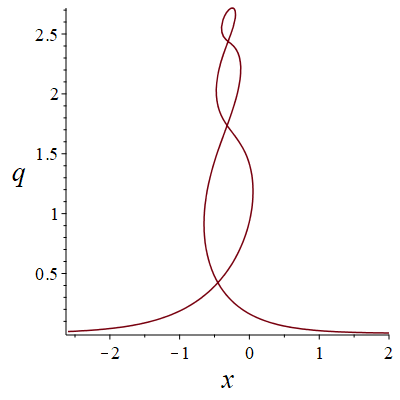}}
	\subfigure[]{\includegraphics[height=4.5cm,width=4.5cm]{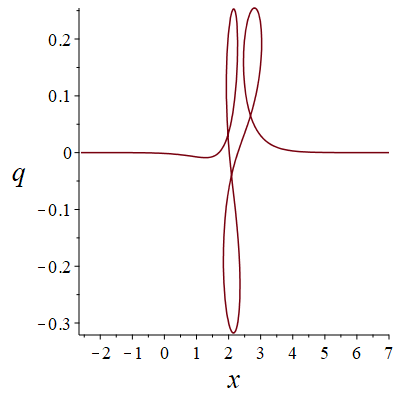}}
	\caption{Profiles of unbounded three-positon solution $q_p^{[3]}$ for the RST nonlocal short pulse equation in $(x,t)-$plane with parameters $\alpha_1=0.4, \alpha_2=-0.5$. Panel (a) is the profile at $t=5$; panel (b) is the profile at $t=0$; panel (c) is the profile at $t=-5$.}
	\label{fig3p3}
\end{figure}

\noindent \textbf{Case 4.2}: When $N=3$, three-positon solutions of the RST nonlocal short pulse equation \eqref{a3} can also be given  by taking  $\lambda_3,\lambda_5\rightarrow \lambda_1$ and $\lambda_4,\lambda_6\rightarrow \lambda_2$. Similar with the case of two-positon, a special case can be considered. When eigenvalues satisfy $\alpha_2=-\alpha_1$, a bounded three-positon can be obtained. The 3D plot of the bounded three-positon are illustrated in Fig.\ref{fig3p1} (a) with the parameters $\alpha_1=0.5,\alpha_2=-0.5$.  Profiles of the bounded three-positon at (a) $t=10$ (b) $t=0$ and (c) $t=-10$ are drawn in Fig.\ref{fig3p2}. As it can be seen in this pictures, when $t\rightarrow -\infty$, the shape and amplitude of three-positon remain unchanged. When $t=0$, three waves colide with each other, the amplitude reaches the maximum value. When $t\rightarrow +\infty$, the shape and amplitude recover as before gradually. Futher, when  $\left| \alpha_1\right| \not =\left| \alpha_2\right| $, the unbounded three-positon can be received. The parameters are choosen as Fig.\ref{fig3p1} (b) with $\alpha_1=0.4,\alpha_2=-0.5$, Fig.\ref{fig3p1}(c) is the plot of $q_p^{[3]}r_p^{[3]}$. the profile of the unbounded three-positon at (a) $t=5$ (b) $t=0$ and (c) $t=-5$ are shown in Fig.\ref{fig3p3}.

\subsection{Mixed solutions of the RST nonlocal short pulse equation}
\begin{figure}[!htbp]
	\centering
	\subfigure[]{\includegraphics[height=4.5cm,width=4.5cm]{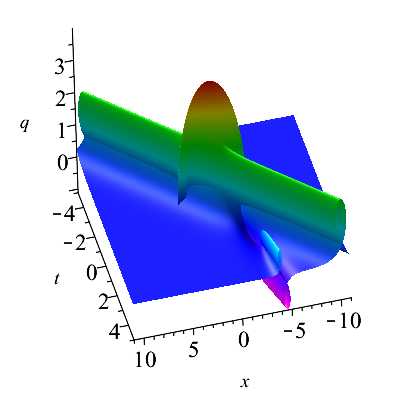}}
	\subfigure[]{\includegraphics[height=4.5cm,width=4.5cm]{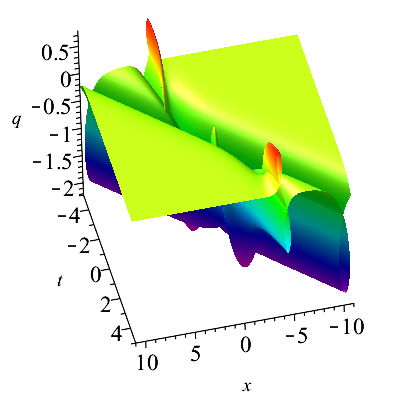}}\\
	\caption{3D plots of the bounded mixed solutions of one-soliton and two-positon $q_{1s,2p}$ for the RST nonlocal short pulse equation in $(x,t)-$plane from case 4.3. Panel (a) is the plot of $q_{1s,2p}$ with parameters $\alpha_1=0.5,\alpha_2=-0.5,\alpha_5=1$ and $\alpha_6=-1$; panel (b) is the plot of $q_{1s,2p}$ with parameters $\alpha_1=0.5,\alpha_2=-0.5,\alpha_5=-1$ and $\alpha_6=1$.}
	\label{fig1s2p1}
	\subfigure[]{\includegraphics[height=4.5cm,width=4.5cm]{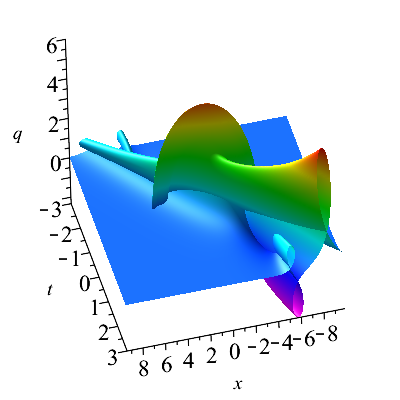}}
	\subfigure[]{\includegraphics[height=4.5cm,width=4.5cm]{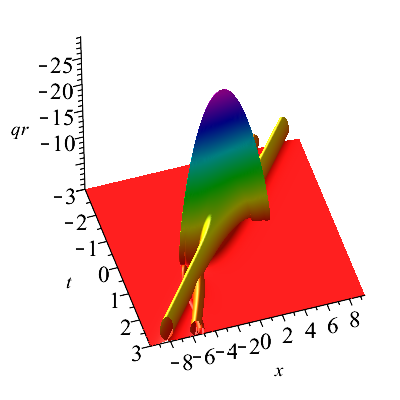}}\\
	\caption{3D plots of the unbounded mixed solutions of one-soliton and two-positon $q_{1s,2p}$ for the RST nonlocal short pulse equation in $(x,t)-$plane with parameters $\alpha_1=0.8,\alpha_2=-0.8,\alpha_5=1$ and $\alpha_6=-1.2$. Panel (a) is the plot of $q_{1s,2p}$; panel (b) is the plot of $q_{1s,2p}r_{1s,2p}$.}
	\label{fig1s2p5}
	\subfigure[]{\includegraphics[height=4.5cm,width=4.5cm]{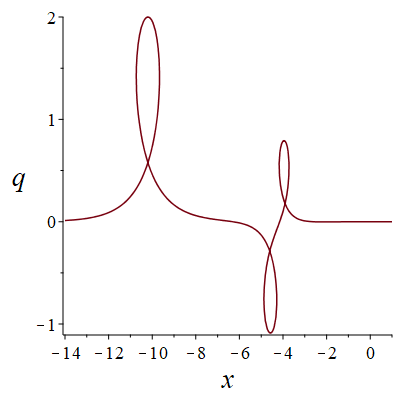}}
	\subfigure[]{\includegraphics[height=4.5cm,width=4.5cm]{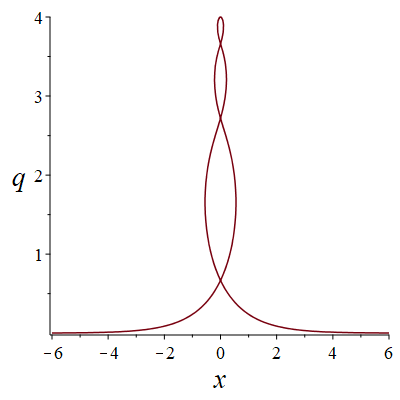}}
	\subfigure[]{\includegraphics[height=4.5cm,width=4.5cm]{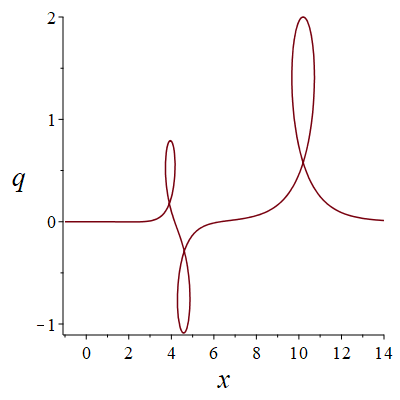}}\\
	\caption{Profiles of bounded mixed solution of one-soliton and two-positon $q_{1s,2p}$ for the RST nonlocal short pulse equation in $(x,t)-$plane with parameters $\alpha_1=0.5, \alpha_2=-0.5,\alpha_5=1, \alpha_6=-1$. Panel (a) is the profile at $t=5$; panel (b) is the profile at $t=0$; panel (c) is the profile at $t=-5$.}
	\label{fig1s2p2}
\end{figure}

\begin{figure}[!htbp]
		\centering
		\subfigure[]{\includegraphics[height=4.5cm,width=4.5cm]{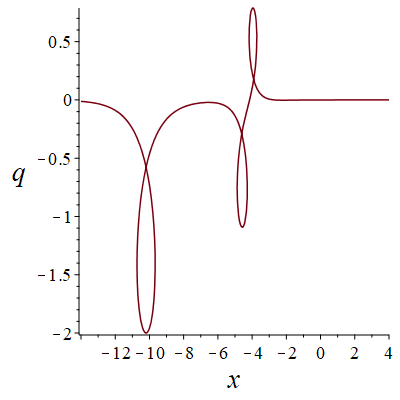}}
	\subfigure[]{\includegraphics[height=4.5cm,width=4.5cm]{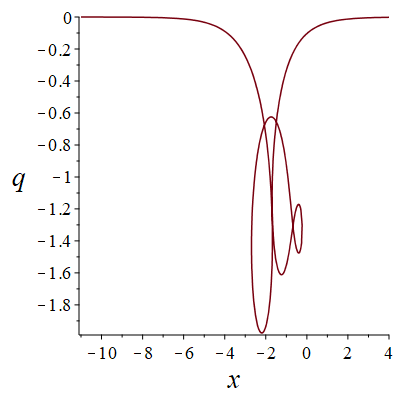}}
	\subfigure[]{\includegraphics[height=4.5cm,width=4.5cm]{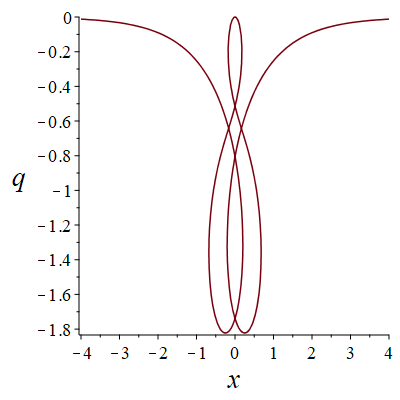}}\\
	\caption{Profiles of bounded mixed solution of one-soliton and two-positon $q_{1s,2p}$ for the RST nonlocal short pulse equation in $(x,t)-$plane with parameters $\alpha_1=0.5, \alpha_2=-0.5,\alpha_5=-1, \alpha_6=1$. Panel (a) is the profile at $t=5$; panel (b) is the profile at $t=1$; panel (c) is the profile at $t=0$.}
	\label{fig1s2p3}
	\subfigure[]{\includegraphics[height=4.5cm,width=4.5cm]{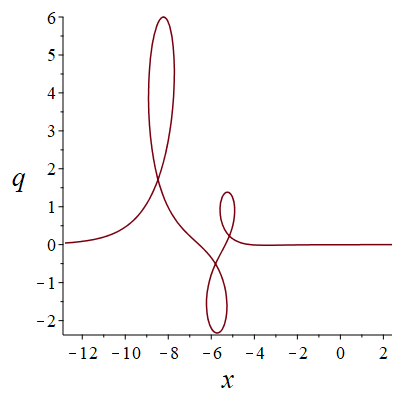}}
	\subfigure[]{\includegraphics[height=4.5cm,width=4.5cm]{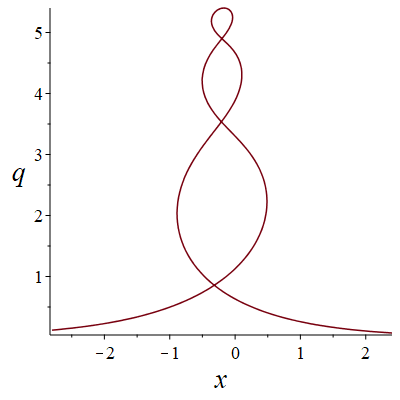}}
	\subfigure[]{\includegraphics[height=4.5cm,width=4.5cm]{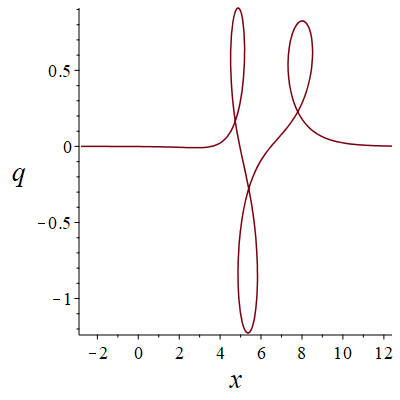}}\\
	\caption{Profiles of unbounded mixed solution of one-soliton and two-positon $q_{1s,2p}$ for the RST nonlocal short pulse equation in $(x,t)-$plane with parameters $\alpha_1=0.8, \alpha_2=-0.8,\alpha_5=1, \alpha_6=-1.2$. Panel (a) is the profile at $t=3$; panel (b) is the profile at $t=0$; panel (c) is the profile at $t=-3$.}
	\label{fig1s2p4}
\end{figure}

In this section, the mixd solution of solitons and positons for the RST nonlocal short pulse equation will be disscussed. The positon solution of RST nonlocal short pulse equation is gained through the higher-order Taylor expansion with $\lambda_{2j-1}\rightarrow \lambda_1, \lambda_{2j}\rightarrow \lambda_2$ in the $N$-soliton solution. However, if some of eigenvalues keep unchanged and others take the limit that $\lambda_{2j-1}\rightarrow \lambda_1, \lambda_{2j}\rightarrow \lambda_2$, then the mixed solutions of solitons and positons can be obtained. Considering following lower order scenarios:

\noindent \textbf{Case 4.3}:When $N=3$, let $\lambda_{3}\rightarrow \lambda_1,\lambda_{4}\rightarrow \lambda_2$ and $\lambda_5, \lambda_6$ maintain the original form. Different kinds of solutions will be generated by taking different eigenvalues. When eigenvalues satisfy $\alpha_2=-\alpha_1,\alpha_6=-\alpha_5$, it yields the bounded mixed solution of one-soliton and two-positon, the plots are shown in Fig.\ref{fig1s2p1} (a)  with the parameters $\alpha_1=0.5,\alpha_2=-0.5, \alpha_5=1,\alpha_6=-1$. Fig.\ref{fig1s2p2} shows the profile of the bounded mixed solution of one-soliton and two-positon at (a) $t=5$ (b) $t=0$ and (c) $t=-5$. It is obvious that the mixed solution are loop in $(x,t)-$plane. In addition, when $\alpha_1=0.5,\alpha_2=-0.5,\alpha_5=-1, \alpha_6=1$, The Figs.\ref{fig1s2p1} (b) and the Figs.\ref{fig1s2p3} show the characteristics of the mixed solution of one-soliton and two-positon, which is also bounded. By analyzing the Fig.\ref{fig1s2p1}, it can be observed that the shape of the solution in Fig.\ref{fig1s2p1} (b) is different from Fig.\ref{fig1s2p1} (a) although they are both bounded. Futhermore, unbounded mixed solution of one-soliton and two-positon can be derived by taking $\alpha_1=-\alpha_2, \left| \alpha_5\right| \not =\left| \alpha_6\right| $, which characteristics are shown in Figs.\ref{fig1s2p5} (a) and Figs. \ref{fig1s2p4}. The shape of $q_{1s,2p}r_{1s,2p}$ is shown in Figs.\ref{fig1s2p5} (b)

\begin{figure}[tbh]
	\centering
	\subfigure[]{\includegraphics[height=4.5cm,width=4.5cm]{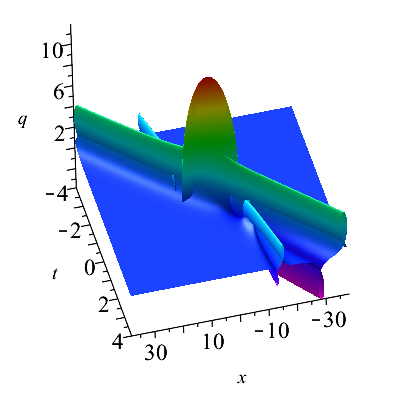}}
	\subfigure[]{\includegraphics[height=4.5cm,width=4.5cm]{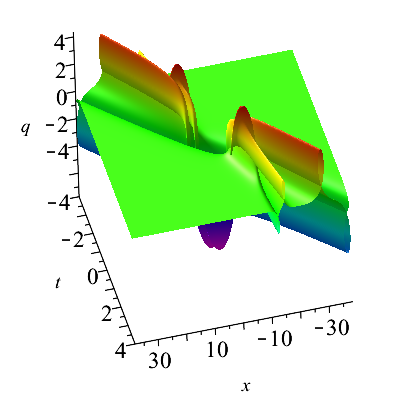}}
	\caption{3D plots of the mixed solutions of two-soliton and two-positon $q_{2s,2p}$ for the RST nonlocal short pulse equation in $(x,t)-$plane from case 4.4. Panel (a) is a bounded mixed solution with parameters $\alpha_1=1,\alpha_2=-1,\alpha_5=2$ and $\alpha_6=-2$; panel (b) is a bounded mixed solution with parameters $\alpha_1=1,\alpha_2=-1,\alpha_5=-2$ and $\alpha_6=2$.}
	\label{fig2s2p1}
	\subfigure[]{\includegraphics[height=4.5cm,width=4.5cm]{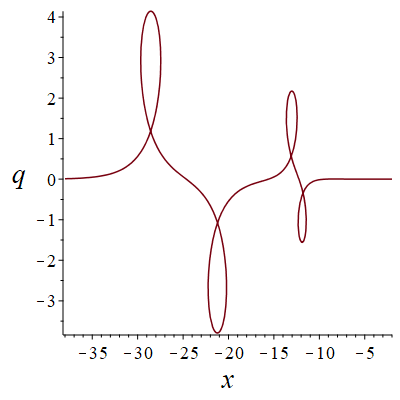}}
	\subfigure[]{\includegraphics[height=4.5cm,width=4.5cm]{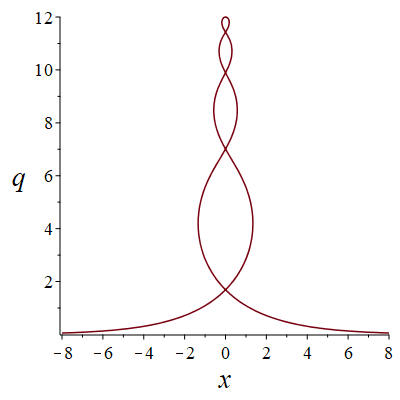}}
	\subfigure[]{\includegraphics[height=4.5cm,width=4.5cm]{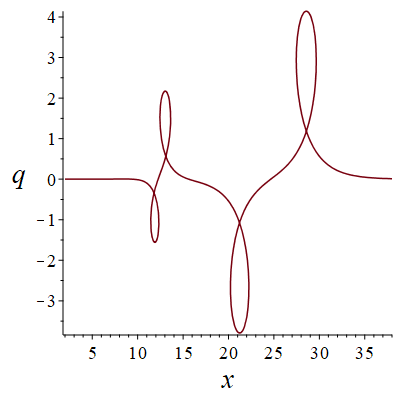}}
	\caption{Profiles of bounded mixed solution of two-soliton and two-positon $q_{2s,2p}$ for the RST nonlocal short pulse equation in $(x,t)-$plane with parameters $\alpha_1=1, \alpha_2=-1,\alpha_5=2, \alpha_6=-2$. Panel (a) is the profile at $t=3$; panel (b) is the profile at $t=0$; panel (c) is the profile at $t=-3$.}
	\label{fig2s2p2}
	\subfigure[]{\includegraphics[height=4.5cm,width=4.5cm]{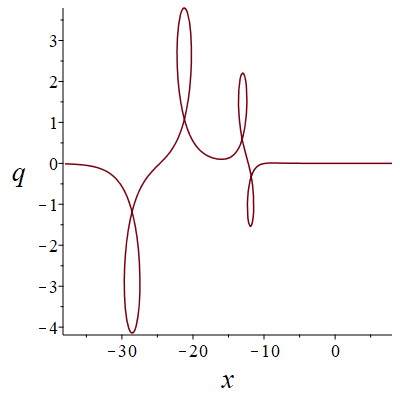}}
	\subfigure[]{\includegraphics[height=4.5cm,width=4.5cm]{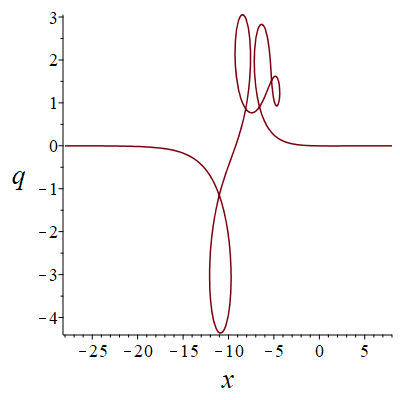}}
	\subfigure[]{\includegraphics[height=4.5cm,width=4.5cm]{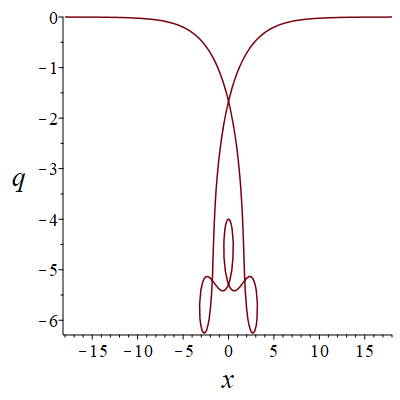}}
	\caption{Profiles of bounded mixed solution of two-soliton and two-positon $q_{2s,2p}$ for the RST nonlocal short pulse equation in $(x,t)-$plane with parameters $\alpha_1=1, \alpha_2=-1,\alpha_5=-2, \alpha_6=2$. Panel (a) is the profile at $t=3$; panel (b) is the profile at $t=1$; panel (c) is the profile at $t=0$.}
	\label{fig2s2p3}
\end{figure}

\begin{figure}[tbh]
	\centering
	\subfigure[]{\includegraphics[height=4.5cm,width=4.5cm]{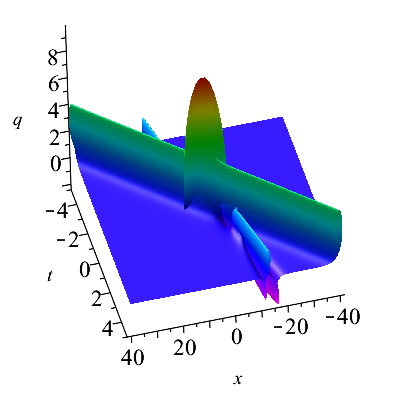}}
	\subfigure[]{\includegraphics[height=4.5cm,width=4.5cm]{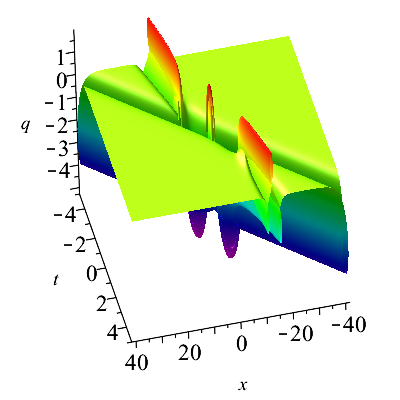}}
	\caption{3D plots of the mixed solutions of two-soliton and two-positon $q_{1s,3p}$ for the RST nonlocal short pulse equation in $(x,t)-$plane from case 4.4. Panel (a) is a bounded mixed solution with parameters $\alpha_1=1,\alpha_2=-1,\alpha_7=2$ and $\alpha_8=-2$; panel (b) is a bounded mixed solution with parameters $\alpha_1=1,\alpha_2=-1,\alpha_7=-2$ and $\alpha_8=2$.}
	\label{fig1s3p1}
	\subfigure[]{\includegraphics[height=4.5cm,width=4.5cm]{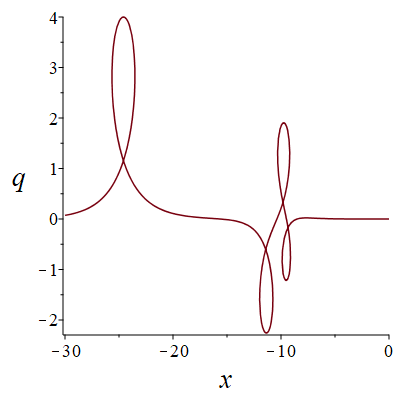}}
	\subfigure[]{\includegraphics[height=4.5cm,width=4.5cm]{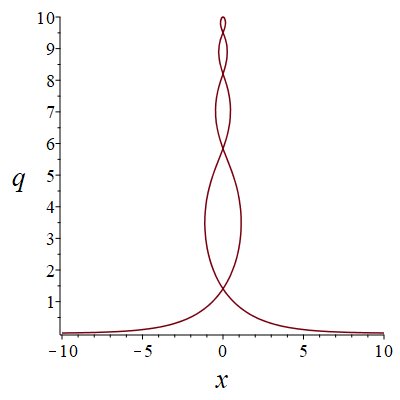}}
	\subfigure[]{\includegraphics[height=4.5cm,width=4.5cm]{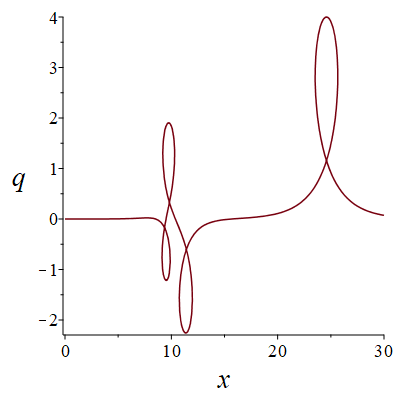}}
	\caption{Profiles of bounded mixed solution of one-soliton and three-positon $q_{1s,3p}$ for the RST nonlocal short pulse equation in $(x,t)-$plane with parameters $\alpha_1=1, \alpha_2=-1,\alpha_5=2, \alpha_6=-2$. Panel (a) is the profile at $t=3$; panel (b) is the profile at $t=0$; panel (c) is the profile at $t=-3$.}
	\label{fig1s3p2}
	\subfigure[]{\includegraphics[height=4.5cm,width=4.5cm]{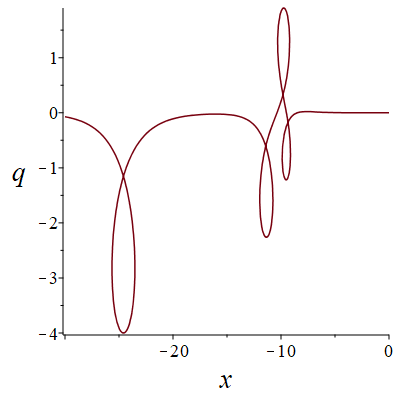}}
	\subfigure[]{\includegraphics[height=4.5cm,width=4.5cm]{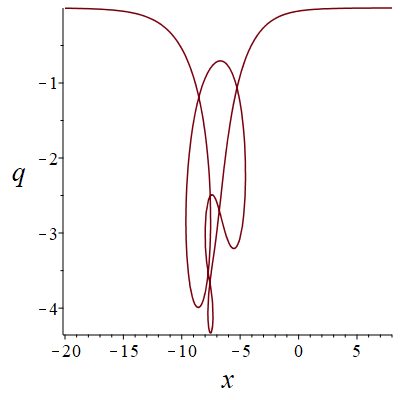}}
	\subfigure[]{\includegraphics[height=4.5cm,width=4.5cm]{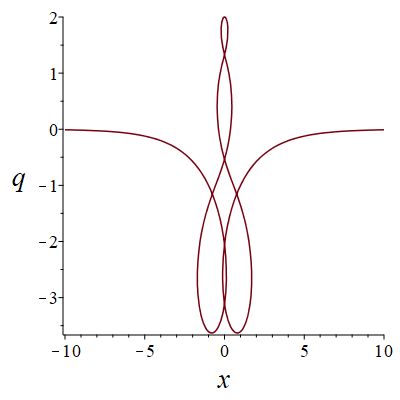}}
	\caption{Profiles of bounded mixed solution of one-soliton and three-positon $q_{1s,3p}$ for the RST nonlocal short pulse equation in $(x,t)-$plane with parameters $\alpha_1=1, \alpha_2=-1,\alpha_5=-2, \alpha_6=2$. Panel (a) is the profile at $t=3$; panel (b) is the profile at $t=1$; panel (c) is the profile at $t=0$.}
	\label{fig1s3p3}
\end{figure}

\noindent \textbf{Case 4.4}:When $N=4$, because it will be more complex to calculate the unbounded mixed solutions, we only consider the bounded mixed solutions. By taking $\lambda_{3}\rightarrow \lambda_1,\lambda_{4}\rightarrow \lambda_2,\lambda_7\rightarrow \lambda_5,\lambda_{8}\rightarrow
\lambda_6, \lambda_2=-\lambda_1,\lambda_6=-\lambda_5$, the bounded mixed solutions of two-soliton and two-positon can be drawn. The suitable parameters are choosen as Fig.\ref{fig2s2p1} (a) with $\alpha_1=1,\alpha_2=-1, \alpha_5=2,\alpha_6=-2$,  Fig.\ref{fig2s2p2} with $t=3,0,-3$. The features of the mixed solutions of two-soliton and two-positon are similar with that of one-soliton and two-positon. The solution shown in Fig.\ref{fig2s2p1}(b) with the parameters $\alpha_1=1,\alpha_2=-1, \alpha_5=-2,\alpha_6=2$, Fig.\ref{fig2s2p3} with $t=3,1,0$. However, when $\lambda_{3}\rightarrow \lambda_1,\lambda_{4}\rightarrow \lambda_2,\lambda_5\rightarrow \lambda_1,\lambda_{6}\rightarrow
\lambda_2, \lambda_2=-\lambda_1,\lambda_8=-\lambda_7$, the mixed solutions of one-soliton and three-positon can be generated. The solution shown in Fig.\ref{fig1s3p1} (a) with the parameters $\alpha_1=1,\alpha_2=-1, \alpha_7=2,\alpha_8=-2$, Fig.\ref{fig1s3p2} with $t=3,0,-3$. The solution shown in Fig.\ref{fig1s3p1}(b) with the parameters
$\alpha_1=1,\alpha_2=-1, \alpha_7=-2,\alpha_8=2$, Fig.\ref{fig1s3p3} with $t=3,1,0$.

For the mixed solutions of positons and solitons, some marked features can be observed from pictures. During the collision, the shapes of soliton and positon keep unchanged. However, the positon have a ``phase shift", while the soliton do not experience any asymptotic phase changes. After the collision, the shape of solitons and positons do not have any change and there exist no ``phase shift". Therefore, the positon is completely transparent.

\section{Conclusions}
In this paper, we studied the determinant representation of the multi-soliton solutions of the RST nonlocal short pulse equation by hodograph transformation and classical Darboux transformation. The loop soliton solutions can be obtained by using classical Darboux transformation. It is worth noting that choosing different eigenvalues can result in different kinds of solutions, which is an interesting phenomenon of nonlocal equation and differ from the previous studies of local equation. Comparing to the local equation, the reduction condition of nonlocal equation occurs some changes. The reduction condition between eigenvectors $\Phi_{2N-1}$ and $\Phi_{2N}$ can be derived in the local equation, which is absent in the nonlocal equation. However,  in both local equation and nonlocal equation, there is reduction condition between eigenfunctions $\phi_j$ and $\psi_j$ corresponding the same eigenvalue $\lambda_j$. Additionally, soliton solutions of the local equation are always bounded. However, two types of solutions can be obtained in nonlocal equation: bounded soliton and unbounded soliton. When eigenvalues are related with a conjugate, the soliton is bounded, which is also the solution of the local equation. When eigenvalues are differently chosen, the soliton with gain or loss is unbounded.  

Furthermore, according to the above conclusions, the analytical expression of the positon solutions of the RST nonlocal short pulse equation is provided by the degenerate DT. The positon solutions for the RST nonlocal short pulse equation is first to be studied. The loop positon can be gained in the $(x,t)-$ plane. It is easy to see that the trajectory of positons is not a straight line. Similar with the soliton solutions discussed above, bounded or unbounded positons will be generated by taking different eigenvalues. Finally, we discuss the mixed solution of 2-positon and 1-soliton, the mixed solution of 3-positon and 1-soliton and the mixed solution of 2-positon and 2-soliton. The positon are not affected by shape changes during their collision and it is completely transparent to soliton.

In addition, there are many meaningful related issues worth further research. For example, the breather solutions have been obtained by nonzero ``seed" solution, it is worth to consider the degenerate breather solutions of the RST nonlocal short pulse equation in the future.

\section{ Acknowledgments}
{This work is supported by
	the Zhejiang Provincial Natural Science Foundation of China under Grant No.LY24A010002, the Natural Science Foundation of Ningbo under Grant No. 2023J126, K.C.Wong Magna Fund in Ningbo University. We thanks the valuable discussions of Professor Jingsong He. }\\

\noindent\textbf{Compliance with ethical standards}\\

\noindent \textbf{Ethical Statement}\\ {\small Authors declare that they comply with ethical standards. } \\

\noindent \textbf{Conflict of interest}\\ {\small Authors declare that they have no conflict of interest.} \\

\noindent\textbf{Data Availability Statement}\\
{\small The data that support the findings of this article are available from the corresponding author, upon reasonable request. } \\

\end{document}